\documentclass[aps,pra,showpacs,twoside,twocolumn,10pt]{revtex4-2}
\usepackage[colorlinks=true, citecolor=red, urlcolor=blue ]{hyperref}
\usepackage{epsfig,newlfont,amssymb,amsfonts,amsmath,bm,subfigure,palatino,mathtools,amsthm,braket,times,soul,enumitem,color}
\usepackage[normalem]{ulem}
\usepackage{xcolor}

\usepackage{blindtext}
\usepackage{graphicx}
\usepackage{amsmath}
\usepackage{bm}
\usepackage{hyperref}
\usepackage{geometry}
\usepackage{amsthm}
\usepackage{gensymb}
\usepackage{physics}
 \definecolor{darkgreen}{rgb}{0.47,0.67,0.19}
\begin{document}

\title{Inhibition of spread of typical bipartite and genuine multiparty entanglement\\in response to disorder}
\author{George Biswas}
\author{Anindya Biswas}
\affiliation{Department of Physics, National Institute of Technology Sikkim, Ravangla, South Sikkim 737 139, India}
\author{Ujjwal Sen}
\affiliation{Harish-Chandra Research Institute, HBNI, Chhatnag Road, Jhunsi, Prayagraj 211 019, India}

%

\begin{abstract}
The distribution of entanglement of typical multiparty quantum states is not uniform over  the range of the measure utilized for quantifying the entanglement. We intend to find the response to disorder in the state parameters on this non-uniformity for typical states. We find that the typical entanglement, averaged over the disorder, is taken farther away from uniformity, as quantified by decreased standard deviation, in comparison to the clean case. The feature is seemingly generic, as we see it for Gaussian and non-Gaussian disorder distributions, for varying strengths of the disorder, and for disorder insertions in one and several state parameters. The non-Gaussian distributions considered are uniform and Cauchy-Lorentz. Two- and three-qubit pure state Haar-uniform generations are considered for the typical state productions. We also consider noisy versions of the initial states produced in the Haar-uniform generations. A genuine multiparty entanglement monotone is considered for the three-qubit case, while concurrence is used to measure two-qubit entanglement.

\end{abstract}
\maketitle
\section{Introduction}

Random numbers have useful applications in classical information theory, including in cryptography, stochastic estimations, etc. Random quantum states and random unitary operators are the quantum analogs of random numbers in quantum information theory~\cite{emerson2003,PhysRevLett.92.187901,Hayden2004,Clowers2008,Ma2016,Russell_2017,Muraleedharan_2019}. 
Moreover, random states occur naturally when a quantum system gets measured in an unknown basis, or when the system is perturbed by an uncontrolled and unknown environment. Similarly, random states can also get generated in the initialization of the corresponding quantum devices. Quantum algorithms and other quantum-enabled protocols~\cite{preskill1999lecture,nielsen_chuang_2010} 
sometime assume that interaction with environment can somehow be avoided. However, unless an error-correction procedure~\cite{preskill1999lecture,nielsen_chuang_2010} 
is incorporated, which can be costly, random states would naturally appear and remain in the corresponding quantum circuits.

 In this paper, we investigate entanglement properties~\cite{RevModPhys.81.865,guhne2009entanglement,Das_2016} 
of randomly generated bipartite and tripartite pure and mixed quantum states, with and without disorder. The case without disorder has, e.g., been  considered in~\cite{PhysRevLett.71.1291,PhysRevLett.72.1148,PhysRevLett.77.1,PhysRevA.74.062314,Dahlsten_2007, Serafini_2007,PhysRevA.84.052321, Muller2012,DeelanCunden2013,Dahlsten_2014,doi:10.1063/1.5119950}. We anticipate a situation where noise from the environment, during the preparation of the state for feeding a quantum circuit or during the evolution of the state through the circuit, gathers as disorder in the parameters of the state written as a superposition over the computational basis. Noise typically acts with a preferred basis, and we assume that basis in our case to be the computational basis, which in turn justifies the building up of disorder in the parameters of the state written in that basis. 
The random bipartite and tripartite states are chosen Haar uniformly. 
The disorder is inferred as quenched, and 
assumed to be spread according to Gaussian, uniform, or Cauchy-Lorentz distributions~\cite{Gentle2012}.

Quantum entanglement is the existence of states of several separated systems that cannot be created by only local quantum operations and classical communication (LOCC) between the systems ~\cite{Einstein1935,schrodinger_1935,PhysRevA.40.4277}. 
There are a large number of measures of entanglement of both bipartite and multipartite quantum states. However, not many of them are computable. We use the concurrence ~\cite{PhysRevLett.78.5022,PhysRevLett.80.2245} 
to measure bipartite entanglement and the computable entanglement monotone  of  Jungnitsch, Moroder, and G{\" u}hne (JMG)~\cite{Otfried} for measuring genuine multiparticle entanglement. 
 We find that the introduction of disorder in multiparty quantum state parameters generically inhibits the spread of average entanglement for typical states of the multiparty systems considered. The considerations are restricted to two- and three-qubit states, of both pure and noisy varieties.



The rest of the paper is arranged as follows.
In Sec.~\ref{II}, we briefly discuss the generation of Haar uniform random states, the probability distributions corresponding to the disorders inserted, and the measures employed to compute quantum entanglement. A short discussion of disorder insertion and its averaging is also given there. Also present there is a brief discussion on skewness and kurtosis of a distribution. In Sec.~\ref{III}, we  discuss the entanglement distributions obtained for bipartite and tripartite Haar-uniformly generated pure states, with and without disorder. Several cases are considered, and are given in separate subsections.
Noisy versions of the input states are also used, that lead to mixed states, and these analyses appear in Secs.~\ref{durbar-dui} and \ref{durbar-chhoi}.
A conclusion is presented 
in Sec.~\ref{IV}. 

\section{Gathering the tools}
\label{II}

\subsection{Haar uniform random states and probability density functions}

The multiparty states utilized for the computation of entanglement and average entanglement have been chosen ``Haar uniformly''. A pure quantum state can be written as 
\begin{equation} \label{qseq}
    \ket{\psi}=\sum_{j=1}^{n}(c_{1j}+ic_{2j})\ket{j}
\end{equation}
where $\ket{j}$ represents the \(j^{\text{th}}\) orthonormal basis vector in the $n$-dimensional Hilbert space, \(\mathbb{C}^n\), and $c_{1j},~c_{2j}$ are real  numbers, 
constrained by the normalization condition, \(\langle \psi | \psi \rangle = 1\).
In particular, for an \(m\)-qubit system, \(n= 2^m\), and \(\mathbb{C}^n = \left(\mathbb{C}^2\right)^{\otimes m}\).
Haar uniformity is attained by choosing the \(c_{ij}\) independently from 
a Gaussian distribution with vanishing mean and 
finite  variance~\cite{Press92numericalrecipes,bengtsson_zyczkowski_2006,cohn2013measure,Dahlsten_2014}. 
The obtained state in each run will have to be normalized to unity.

The probability density function for the Gaussian distribution is given by 
\begin{equation} \label{e2}
f_G(x)=\frac{1}{\sigma_G\sqrt{2\pi}}e^{-\frac{1}{2}\left(\frac{x-\mu_G}{\sigma_G}\right)^2},
\end{equation}
where $\mu_G$ is the mean and $\sigma_G$ is the standard deviation of the distribution. The corresponding semi-interquartile range, which is half of the difference between the 
third and first quartiles
of the distribution function, is given by 
\begin{equation}
\gamma_G\approx0.67448\times\sigma_G.
\end{equation}
%
%

\begin{figure}[htpb]
\centering

\textbf{\textcolor{red}{(a)}}
\includegraphics[width=6cm,height=6cm]{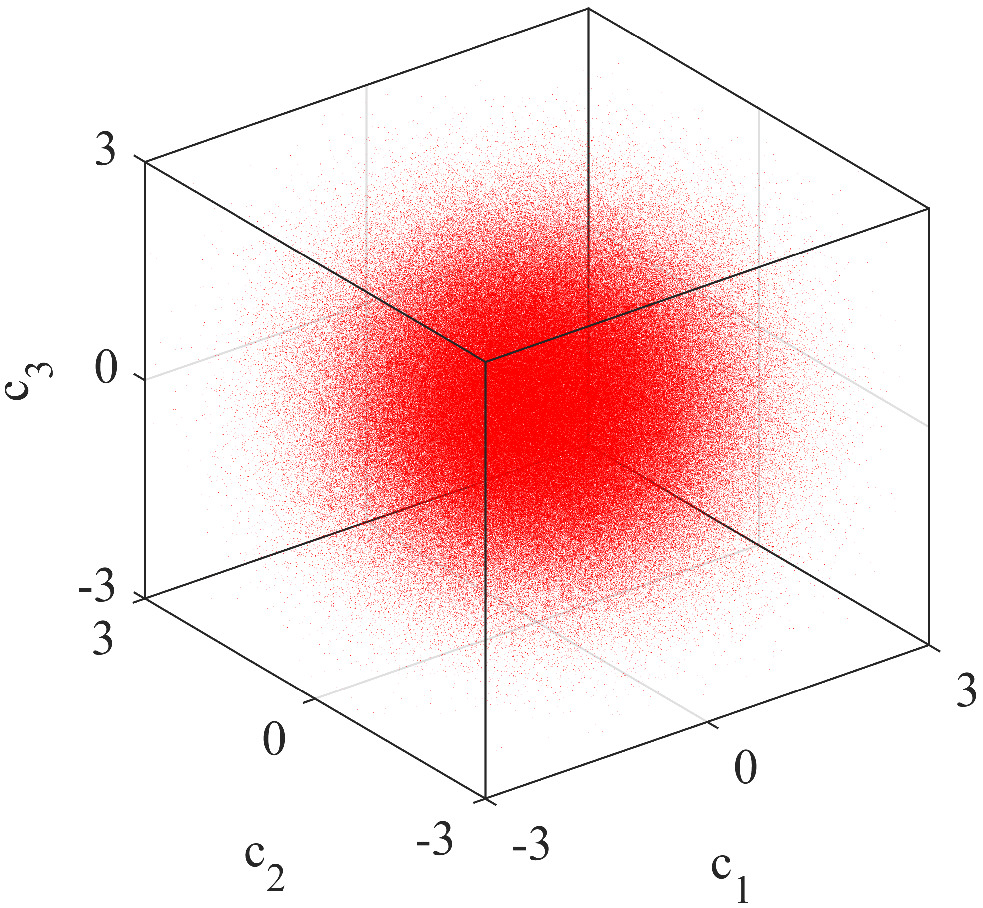}

\textbf{\textcolor{darkgreen}{(b)}}
\includegraphics[width=6cm,height=6cm]{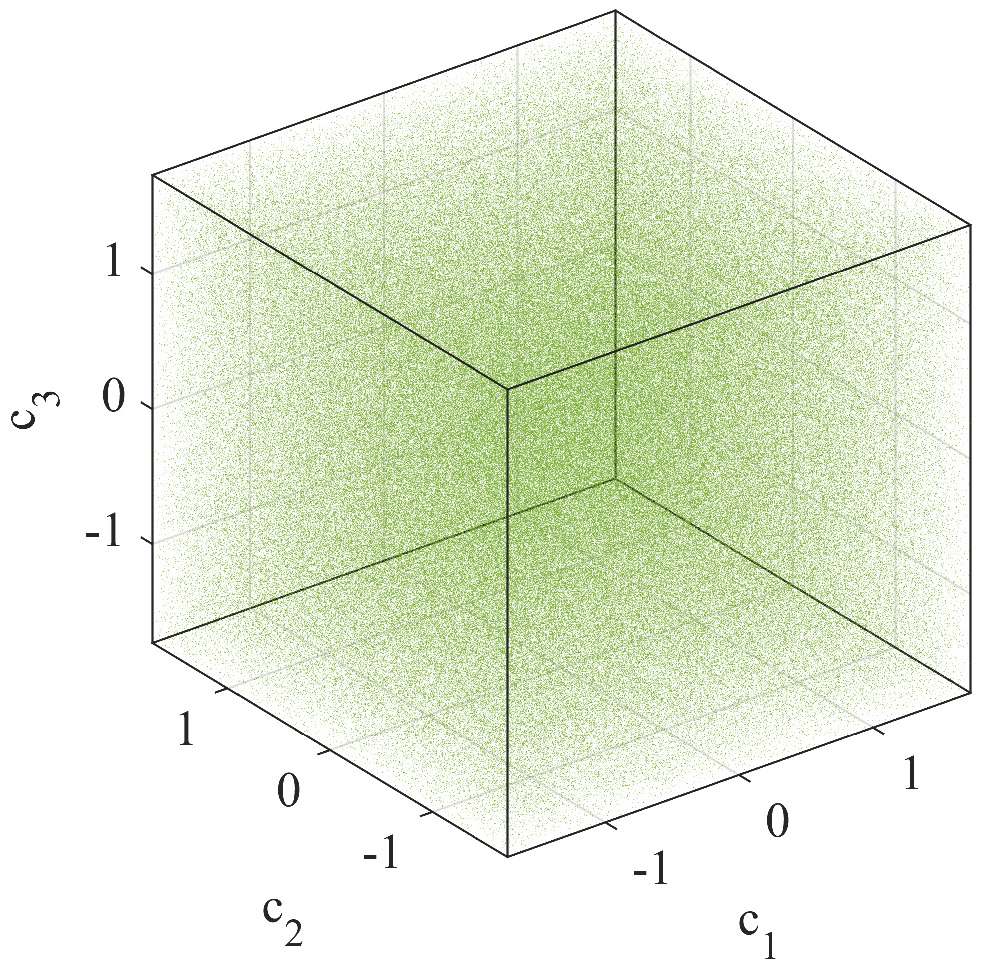}

\textbf{\textcolor{blue}{(c)}}
\includegraphics[width=6cm,height=6cm]{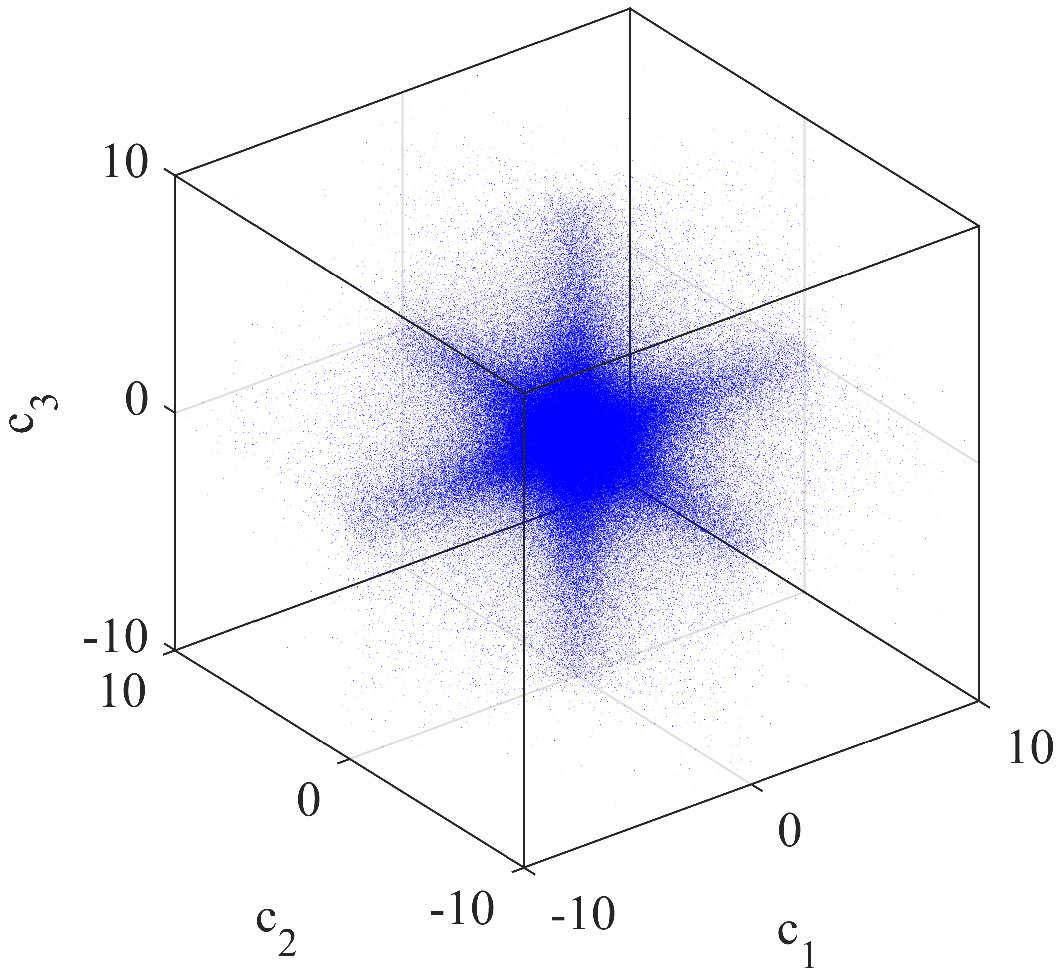}
\caption{Haar-uniform quantum state generation and comparison with other random generations: un-normalized case. The scatter diagrams in the different panels are generated by independently choosing \(c_1\), \(c_2\), \(c_3\) of Eq.~\ref{ieq}, from different probability distributions. The scatter diagram in panel \textcolor{red}{(a)} corresponds to Haar-uniform state generation in a \emph{real} three-dimensional Hilbert space, where the state is not normalized. It is generated by choosing the \(c_i\) independently from the Gaussian distribution 
with mean $\mu_G=0$ and standard deviation $\sigma_G=1$. In panel \textcolor{darkgreen}{(b)}, the \(c_i\) are independently chosen from the uniform distribution with mean $\mu_U=0$ and standard deviation $\sigma_U=1$, so that 
the range is $-\sqrt{3}$ to $\sqrt{3}$, while in panel \textcolor{blue}{(c)}, we use the Cauchy-Lorentz distribution of median $x_0=0$ and semi-interquartile range $\gamma_{C-L} \approx 0.67448$. Note that the Gaussian distribution with standard deviation $\sigma_G=1$ has a semi-interquartile range of $\gamma_G \approx 0.67448$. Note also that the states generated in panels \textcolor{darkgreen}{(b)} and \textcolor{blue}{(c)} are \emph{not} Haar-uniform. All quantities plotted are dimensionless.}
\label{hfig1}
\end{figure}
\begin{figure}[htpb]
\centering

\textbf{\textcolor{red}{(a)}}
\includegraphics[width=7.6cm]{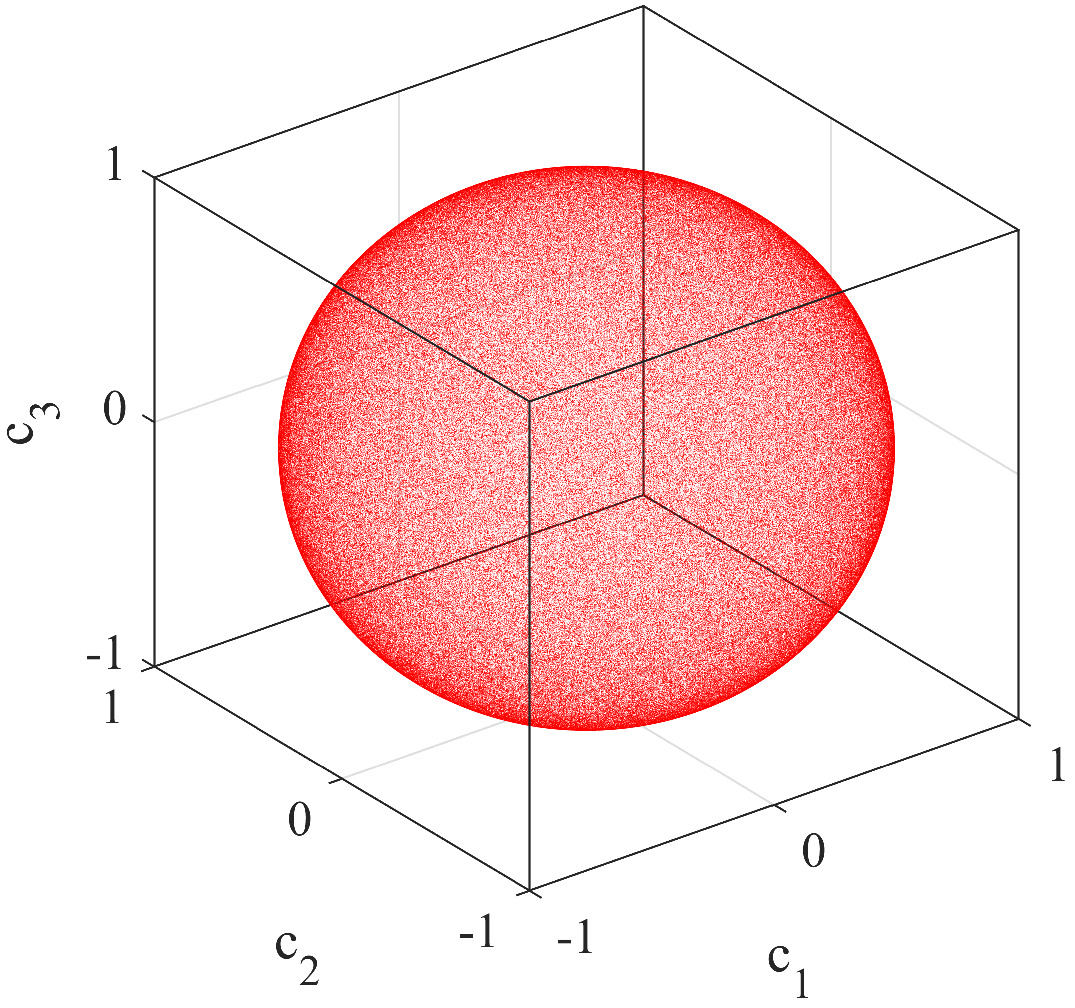}

\textbf{\textcolor{darkgreen}{(b)}}
\includegraphics[width=7.6cm]{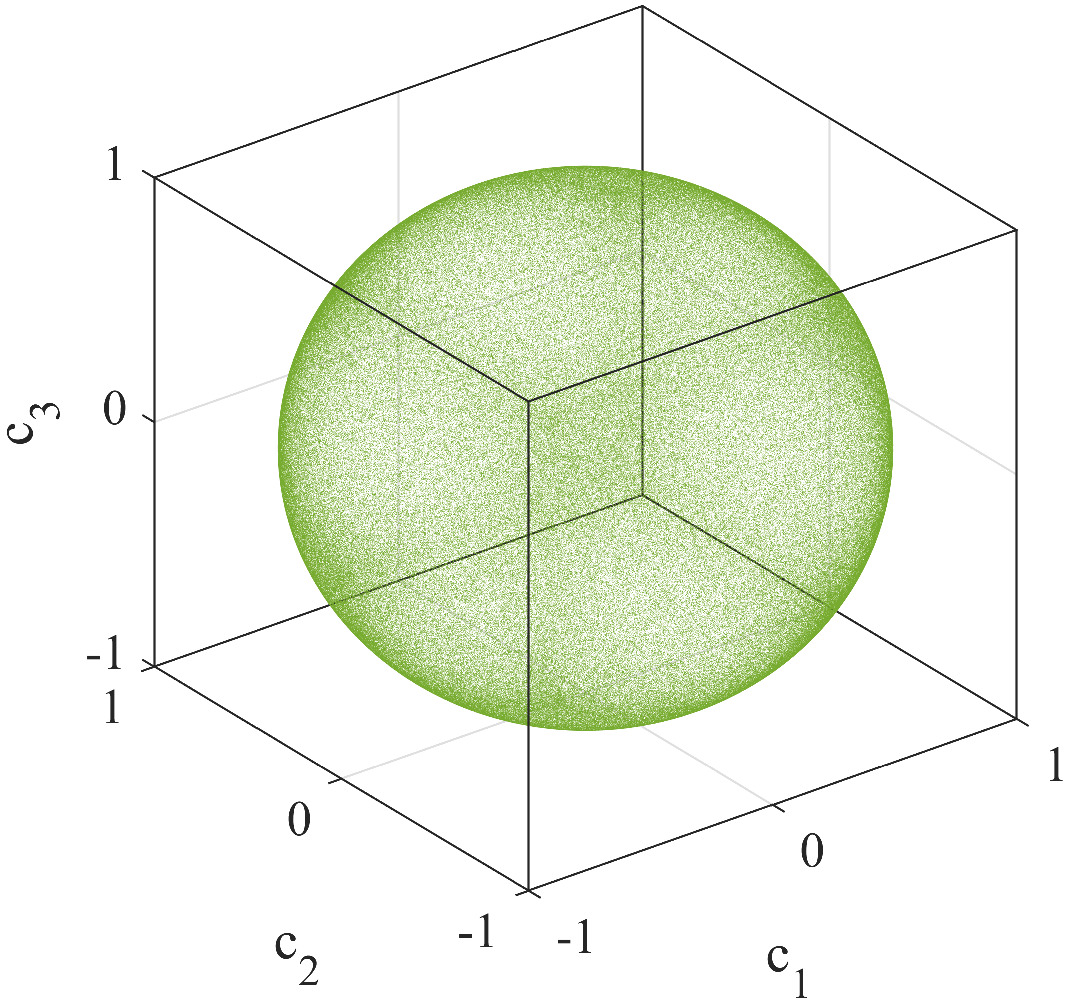}

\textbf{\textcolor{blue}{(c)}}
\includegraphics[width=7.6cm]{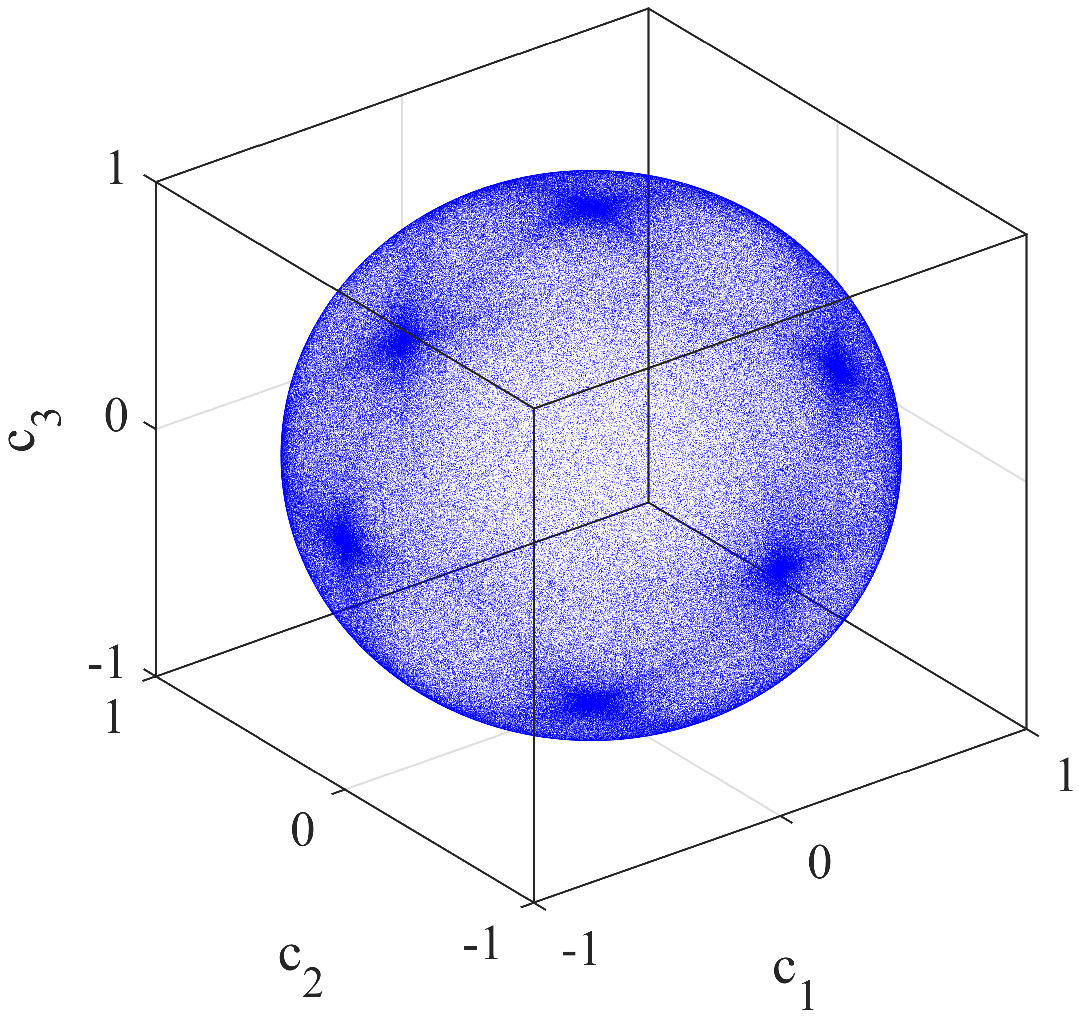}
\caption{Haar-uniform quantum state generation and comparison with other random generations: normalized case. The considerations in the scatter diagrams here are exactly the same as in the respective panels in Fig.~\ref{hfig2}, except that the quantum states are normalized to unity here. 
}
\label{hfig2}
\end{figure}

The un-normalized random states are distributed in a 
hyperspace
of dimension $n$, when the real numbers are chosen from the Gaussian distribution with zero mean and unit variance. Once the states are normalized, they are distributed uniformly over a hypersphere of unit radius in that space. As an illustration, let us consider the space of three orthonormal vectors with \emph{real} coefficients, 
\begin{equation} \label{ieq}
    \ket{\phi}=\sum_{i=1}^{3}c_i\ket{i}.
\end{equation}
We can depict this state as the point \((c_1, c_2, c_3)\) in \(\mathbb{R}^3\).  
%
In Fig.~\ref{hfig1}(a), we plot a scatter diagram of these points for a large number of realizations of the un-normalized state, \(|\phi\rangle\). 
It may be noted that the joint probability distribution of three independent variables, \(x\), \(y\), \(z\), is spherically symmetric,  
if  the individual distributions are Gaussian. For example, with mean $\mu_G=0$ and standard deviation $\sigma_G=1$, the joint probability distribution of \(x\), \(y\), \(z\) is
\begin{equation} \label{jpdeq}
    f_G(x)f_G(y)f_G(z)=(2\pi)^{-\frac{3}{2}}e^{-\frac{x^2+y^2+z^2}{2}}.
\end{equation}
The effect of normalization of the state can be seen in 
Fig.~\ref{hfig2}(a). 
For comparison, we also plot, in the other panels in Figs.~\ref{hfig1} and~\ref{hfig2},  the un-normalized and normalized states, when \(c_1\), \(c_2\), and \(c_3\) are chosen, independently, from two other distributions separately. 


\par The random numbers utilized to model the disorder, arising in the coefficients of randomly chosen bipartite and tripartite quantum states expressed in the computational basis,  
have been chosen from Gaussian, uniform, or Cauchy-Lorentz probability distributions. 

\par The probability density function for the uniform distribution is
\begin{equation} \label{e3}
    f_U(x)= 
\begin{cases}
    \frac{1}{b-a},& \text{if } a\leq x\leq b,\\
    0,              & \text{otherwise.}
\end{cases}
\end{equation}
The mean, standard deviation and semi-interquartile range in this case are given respectively by 
\begin{equation}
\mu_U=\frac{b+a}{2} \quad \text{,} \quad \sigma_U=\frac{b-a}{2\sqrt{3}} \quad \text{and} \quad \gamma_U=\frac{b-a}{4}.
\end{equation}
The probability density function of the uniform distribution in terms of its 
mean and standard deviation 
can be written as
\begin{equation} \label{e5}
    f_U(x)= 
\begin{cases}
    \frac{1}{2\sqrt{3}\sigma_U},& \text{if } \mu_U-\sqrt{3}\sigma_U\leq x\leq \mu_U+\sqrt{3}\sigma_U,\\
    0,              & \text{otherwise.}
\end{cases}
\end{equation}
\par The Cauchy-Lorentz probability density function 
is given by
\begin{equation} \label{e7}
f_{C-L}(x|x_0,\gamma_{C-L})=\frac{\gamma_{C-L}}{\pi\left[\gamma_{C-L}^2+\left(x-x_0\right)^2\right]},
\end{equation}
where $x_0$ is the median of the distribution and $\gamma_{C-L}$ is the scale parameter, being equal to its half width at half maximum or semi-interquartile range. The mean and variance of the Cauchy-Lorentz distribution are not well-defined. The Cauchy principal value of the mean does exist, and equals the median. In absence of the mean, we use the median as a measure of central tendency of the distribution. And in absence of the standard deviation, we use the semi-interquartile range as a measure of dispersion of the distribution. The corresponding  cumulative distribution function 
is 
\begin{align} \label{e8}
F_{C-L}(x|x_0,\gamma_{C-L})&=\int_{-\infty}^{x}f_{C-L}(x'|x_0,\gamma_{C-L})\dd{x'} \nonumber \\
&=\frac{1}{\pi}{\tan}^{-1}\left(\frac{x-x_0}{\gamma_{C-L}}\right)+\frac{1}{2}.
\end{align}
Therefore, the quantile function or the inverse cumulative distribution function is
\begin{equation} \label{e9}
x=x_0+\gamma_{C-L}\tan\left[\pi\left(F_{C-L}-\frac{1}{2}\right)\right].
\end{equation}
This quantile function generates a random number from the Cauchy-Lorentz distribution when $F_{C-L}$ is randomly chosen from a uniform distribution in the range 0 to 1.

\subsection{Entanglement measures}

We wish to analyze entanglement in  bipartite and  tripartite quantum states. The bipartite quantum states that we will encounter are all two-qubit states, and therefore, we can use the concurrence to measure their entanglement contents. 
Concurrence of a two-qubit density matrix $\rho$ is defined as 
\begin{equation} \label{e1}
C(\rho)=\max\{0,\lambda_1-\lambda_2-\lambda_3-\lambda_4\},
\end{equation}
where the $\lambda_i\text{'s}$ are  square roots of the eigenvalues of $\rho\tilde{\rho}$ in descending order. Here $\tilde{\rho}$ is the spin-flipped $\rho$ : $\tilde{\rho}=(\sigma_y\otimes\sigma_y)\rho^*(\sigma_y\otimes\sigma_y)$, $\rho^*$ is the complex conjugate of $\rho$ in the computational basis~\cite{PhysRevLett.78.5022,PhysRevLett.80.2245}. 
The physical meaning of the concurrence is obtained through its relation, for two-qubit states, with the entanglement of formation~\cite{PhysRevA.53.2046,PhysRevA.54.3824}. 
And the entanglement of formation of a bipartite state, not necessarily of two qubits, is a quantifier of the ``amount'' of singlets necessary to create the state by LOCC.

We now move over to the tripartite case, where we use a computable multiparty entanglement monotone, for both pure and mixed states, given in Ref.~\cite{Otfried}.  
The measure is equal to the negativity~\cite{Peres,HORODECKI19961,PhysRevA.58.883,doi:10.1080/09500340008235138,PhysRevA.65.032314,PhysRevLett.95.090503}
for the bipartite case, and can be considered to be an extension of negativity to the multipartite case.
A tripartite state $\rho_{\textnormal{ABC}}$ is not  biseparable (i.e., not a convex combination of states which are separable in at least one bipartition) and therefore genuinely multiparty entangled if
\begin{equation}\label{eqn-gte}
 \rho_{\textnormal{ABC}}\ne p_1\rho^{\textnormal{sep}}_{\textnormal{A}|\textnormal{BC}}+p_2\rho^{\textnormal{sep}}_{\textnormal{B}|\textnormal{CA}}+p_3\rho^{\textnormal{sep}}_{\textnormal{C}|\textnormal{AB}},
\end{equation}
where $\rho_{\textnormal{A}|\textnormal{BC}}^{\textnormal{sep}}=\sum_kq_k\ket{\phi^k_{\textnormal{A}}}\bra{\phi^k_{\textnormal{A}}}\otimes\ket{\psi^k_{\textnormal{BC}}}\bra{\psi^k_{\textnormal{BC}}}$, etc., and $\{p_i\}$, $\{q_k\}$ are probability distributions. Since any biseparable state is a PPT mixture (i.e., is a convex combination of states that are non-negative under partial transpose (PPT) in at least one bipartition), a state which is not a PPT mixture is necessarily genuine multipartite entangled~\cite{Peres,HORODECKI19961}. Therefore, the state $\rho_{\textnormal{ABC}}$ is genuinely multiparty entangled if
\begin{equation}\label{eqn-gte}
 \rho_{\textnormal{ABC}}\ne p_1\rho^{\textnormal{ppt}}_{\textnormal{A}|\textnormal{BC}}+p_2\rho^{\textnormal{ppt}}_{\textnormal{B}|\textnormal{CA}}+p_3\rho^{\textnormal{ppt}}_{\textnormal{C}|\textnormal{AB}},
\end{equation}
where $\rho^{\textnormal{ppt}}_{\textnormal{A}|\textnormal{BC}}$, etc. are states which have a non-negative partial transpose with respect to the indicated bipartition. The genuine  entanglement content of this state is characterized and quantified by an entanglement witness $W$, which is an observable that is non-negative on all biseparable states but has a negative expectation value on at least one entangled state~\cite{Dariusz}. A witness $W$ is fully decomposable if, for every subset $M$ of a system, it is decomposable with respect to the bipartition given by $M$ and its complement $\overline{M}$, which implies that there exist positive semidefinite operators $P_M$ and $Q_M$ 
such that
\begin{equation}
 \textnormal{for all }M: W=P_M+Q_M^{T_M},
\end{equation}
where $T_M$ is the partial transpose with respect to $M$. This observable is non-negative 
on all PPT mixtures, as it is 
so
on all states which are PPT with respect to some bipartition. 
\par Given a multipartite state $\rho$, if \begin{equation}
\min\text{Tr}(W\rho)
\end{equation}
is negative,
then $\rho$ is not a PPT mixture, so that it is genuinely multipartite entangled. The negative of the witness expectation value, with the conditions $0\le P_M\le 1$ and $0\le Q_M\le 1$, is defined as the measure of genuine multipartite entanglement. It satisfies the properties of a good entanglement measure~\cite{Otfried}.
\par The matlab code for this multipartite entanglement measure which require semi-definite programming~\cite{Lieven} has been provided by B. Jungnitsch at~\cite{Bastian}, and has been used in the calculation of the measure 
in this work. 

\subsection{ Disorder and averaging}

Disorder can appear in different hues and patterns in a physical system. We consider here a type of disorder that has often been referred to in the literature as ``quenched''~\cite{doi:10.1142/0223,doi:10.1142/0271,Chakrabarti1996,nishimori01,sachdev_2011,Suzuki2013}. 
This is to be differentiated from the quenching in the dynamics of a physical system. Within this type of disorder, which is also called ``glassy'', the equilibration of the disorder in the system takes a time that is several orders of magnitude higher than the time required to observe the system characteristics which are relevant to our purposes. The system's physical characteristics, as understood from the values of its physical quantities under consideration, have to be averaged over the disorder to obtain physically meaningful numbers. However, because of the nature of disorder chosen, the averaging needs to be performed only after all physical quantities for given realizations of the disorder have been already calculated. This mode of averaging has sometimes been  referred to in the literature as ``quenched averaging''~\cite{Saha_1994,PhysRevE.72.061905,Blavatska_2013}. 
We will however refer to the disorder considered and its averaging without any adjectives.

\subsection{Third and fourth moments}

For a data set with data points $\{(x_1,y_1), (x_2,y_2), ..., (x_N,y_N)\}$ the skewness and kurtosis are defined as follows:
\begin{equation} \label{new_eq_1}
s=\dfrac{\sum_{i=1}^{N}{(y_i-\mu)^3}}{N \sigma^3},
\end{equation}
\begin{equation} \label{new_eq_2}
k=\dfrac{\sum_{i=1}^{N}{(y_i-\mu)^4}}{N \sigma^4},
\end{equation}
where $\mu$ and $\sigma$ are the mean and the standard deviation of the corresponding data set. Skewness is a measure of asymmetry of the distribution. It is negative when the distribution is left skewed or has a longer left tail and positive for a right skewed distribution or a distribution with a longer right tail~\cite{Hippel2011}. 
Kurtosis is a measure which determines the tendency of the distribution to produce outliers~\cite{Westfall2014}. Note that up to the scaling by a power of the standard deviation, the skewness and kurtosis are respectively the third and fourth central moments.

\section{Response of bipartite and tripartite entanglement to disorder}
\label{III}
\par An arbitrary two-qubit pure state is given by 
\begin{equation} \label{e10}
\begin{split}
\left|\psi\right\rangle=&(\alpha_1+{i\alpha}_2)\left|00\right\rangle+(\beta_1+{i\beta}_2)\left|01\right\rangle\\
&+(\gamma_1+{i\gamma}_2)\left| 10\right\rangle+(\delta_1+{i\delta}_2)\left| 11\right\rangle,
\end{split}
\end{equation}
 where 
 \(\alpha_1\), \(\alpha_2\), \(\beta_1\), 
 \(\beta_2\), \(\gamma_1\), \(\gamma_2\), \(\delta_1\), \(\delta_2\)
 are real numbers, constrained by the normalization condition, \(\langle \psi|\psi\rangle =1\). 
 The state in 
 Eq.~(\ref{e10}) can be  Haar-uniformly generated by randomly choosing the eight real coefficients  independently from  a normal (Gaussian) distribution of mean $\mu_G = 0$ and a finite standard deviation, 
 followed by a  normalization~\cite{MISZCZAK2012118,doi:10.1063/1.3595693,e20100745,PhysRevA.93.032125,Dahlsten_2014,cohn2013measure,Zyczkowski_2001}. Here we choose the random numbers  independently from the standard normal distribution, i.e., the Gaussian distribution with mean $\mu_G = 0$ and standard deviation $\sigma_G = 1$.

\begin{figure}[h]
  \centering
    \includegraphics[width=\linewidth]{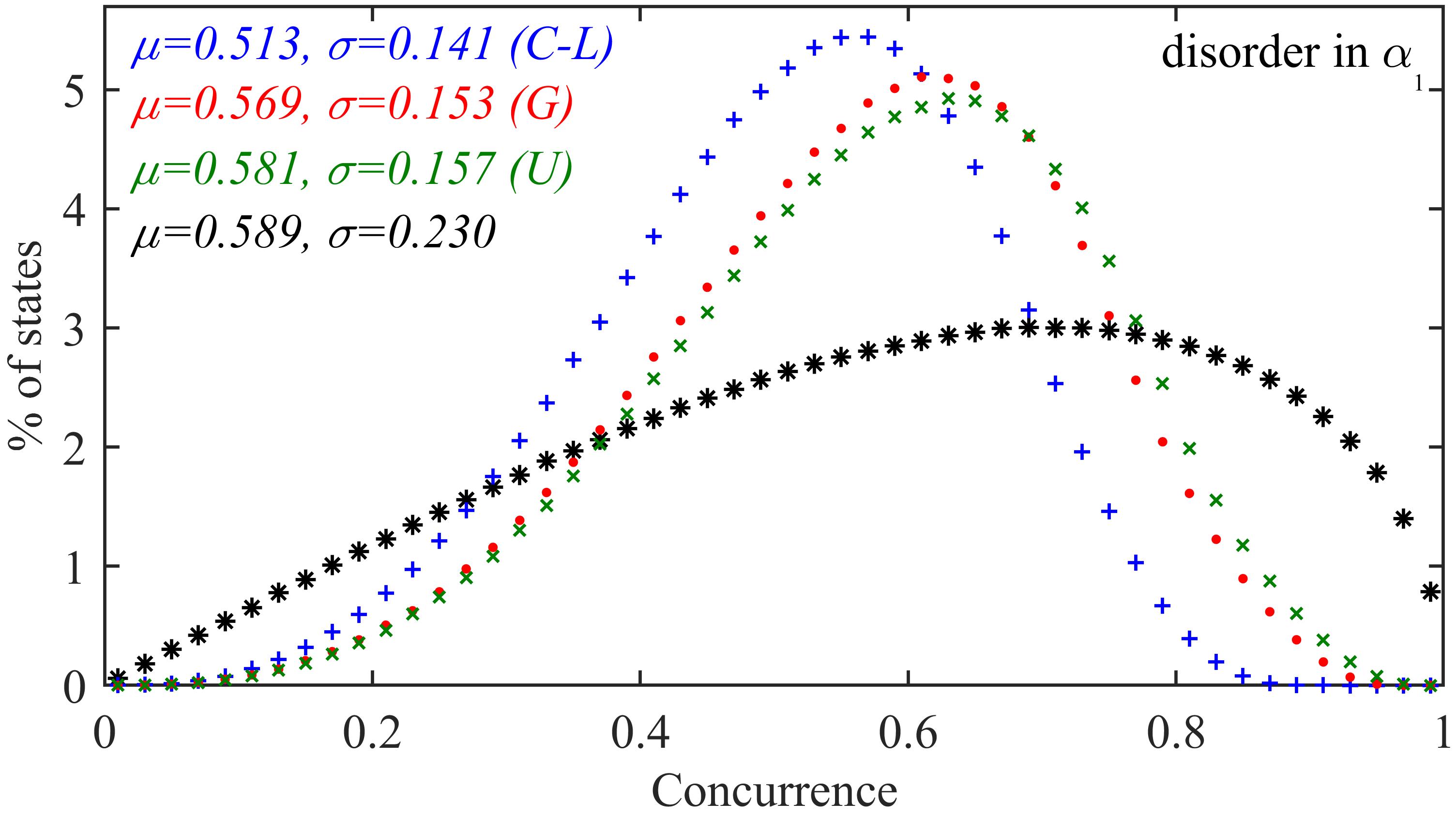}
    \caption{Inhibition of spread of two-party pure quantum state entanglement in response to disorder. We plot here the percentages (precisely, relative frequency percentages) of Haar-uniformly chosen two-qubit pure states \(|\psi\rangle\), given by Eq.~(\ref{e10}), with and without disorder at $\alpha_1$, against the concurrence values, with the latter ranging between $0$ and $1$. The black asterisks correspond to percentages of random two-qubit pure states generated Haar uniformly, while the other three curves depict entanglement distributions for random two-qubit pure states with disorder at $\alpha_1$ chosen from Gaussian (\(G\), red dots), uniform (\(U\), green crosses), and Cauchy-Lorentz (\(C-L\), blue pluses) distributions. The 
    disorder averaged values plotted in the figure are for \(100\) disorder configurations for every \(\alpha_1\). The figure does not change up to the precision used for averaging over \(50\) configurations. The initial Haar-uniform generation (in the ``first step'') used for the figure utilizes \(10^7\) states (and hence the same number of \(\alpha_1\)s), although the same plot does not alter, up to the precision used, for \(10^6\) points. The precision is checked to three significant figures. The calculation of the relative frequency percentages requires a window on the horizontal axis, and it is 0.02 ebits. The convergence checking 
    is done for all other later figures also, but is not explicitly repeated there. The numbers corresponding to the Haar-uniform generation and disorder generation remain the same throughout except the last two figures depicting three qubit entanglement. The precision and the horizontal axis window are also the same in all figures, except the last two, which relate to three-qubit systems. The vertical axis is dimensionless, while the horizontal one is in ebits, which again is the same in all figures except the last two. The skewness and kurtosis of the plot in the ordered case are respectively $s=-0.292$ and $k=2.20$, while for the disordered cases,  they  
    are $s(U)=-0.337$, $k(U)=2.71$; $s(G)=-0.345$, $k(G)=2.73$; $s(C-L)=-0.318$, $k(C-L)=2.68$, respectively for disorders from uniform, Gaussian, and Cauchy-Lorentz distributions.}
  \label{f1}
\end{figure}

\noindent ``\underline{First step.}'' A large number ($10^7$) of such random pure states are generated and the concurrence of each state is calculated. Thus, an entanglement distribution for random two-qubit pure states 
is obtained, with the range being from 0 till 1. The relative frequency percentages of the distribution for different concurrences is plotted as black asterisks in Fig.~\ref{f1}.

\noindent ``\underline{Second step.}'' Next, we introduce disorder at $\alpha_1$ using  Gaussian, uniform, or  Cauchy-Lorentz distribution functions. We begin with the Gaussian case. One hundred random pure states are generated by choosing random numbers from the Gaussian distribution with $\mu_G = \alpha_1$, where $\alpha_1$ is the random number generated in the first step {\textcolor{black}{(after normalization)}}, and $\gamma_G = 1/2$, which corresponds to $\sigma_G \approx 0.74131$. The value of $\sigma_G$ here should not be confused with the value of $\sigma_G$ in the context of Haar uniformity. The value of $\sigma_G$ here refers to the scope of error in the chosen coefficient only. These random numbers are the new $\alpha_1$s (in the disordered case), while the other random numbers are the ones selected in the first step, 
\textcolor{black}{post normalization}. The random pure states are normalized and their average entanglement calculated.
This therefore provides us with another set of \(10^7\) 
disorder averaged entanglement values, and again for this distribution, we plot the corresponding relative frequency percentages for different concurrences, as red dots in Fig.~\ref{f1}. 

The average entanglement for the Haar-uniformly chosen random pure two qubit states is $0.589$, while the corresponding standard deviation 
is $0.230$. This is the clean case, i.e., the case without disorder. 
It can be seen from Fig.~\ref{f1} that introduction of disorder in the parameter $\alpha_1$ from Gaussian  distribution \textcolor{black}{slightly reduces the average entanglement of the states} 
(equalling \(0.569\)). However, the standard deviations of the 
\textcolor{black}{disorder induced}  entanglement distributions are \textcolor{black}{significantly} reduced (being \(0.153\)).

We therefore find that the distribution of entanglement of Haar-uniformly chosen quantum states is not uniform in the entire range of possible values, viz. \([0,1]\), but is concentrated at an intermediate point in \([0,1]\). Moreover, \emph{introduction of disorder in the parameters of the Haar-uniformly generated quantum states leads to a further concentration of values around an intermediate central value}. This of course has been obtained in a case that is highly specific from several perspectives. \emph{The question that we ask is whether this feature is generic.} To answer this question, we try to remove the specificity of the case studied, viz. checking the response in pure two-party quantum entanglement to Gaussian disorder in a single parameter, from several perspectives.  

The Haar-uniform generation of quantum states spreads out the states in the most uniform pattern on the state space. Any variation of that by putting in noise or disorder will make it less uniform. However, this does not necessarily mean that the less uniform distribution has lower standard deviation, as the lower uniformity can result in clustering at different places in the state space. As an example, let us consider 101 points uniformly distributed over the range \([0,1]\). This distribution has a standard deviation of approximately \(0.29\). However, if the same points are non-uniformly distributed with 51 at 0 and 50 at 1, the standard deviation will be approximately \(0.50\). Moreover, we calculate the standard deviations of the distributions of entanglement of the states obtained from a given distribution with or without disorder insertion. Entanglement is a nonlinear function of the state parameters, which makes it even more  nontrivial to easily infer the dispersion at the level of entanglement from that at the level of state parameters. 

\begin{figure}[h]
  \centering
    \includegraphics[width=\linewidth]{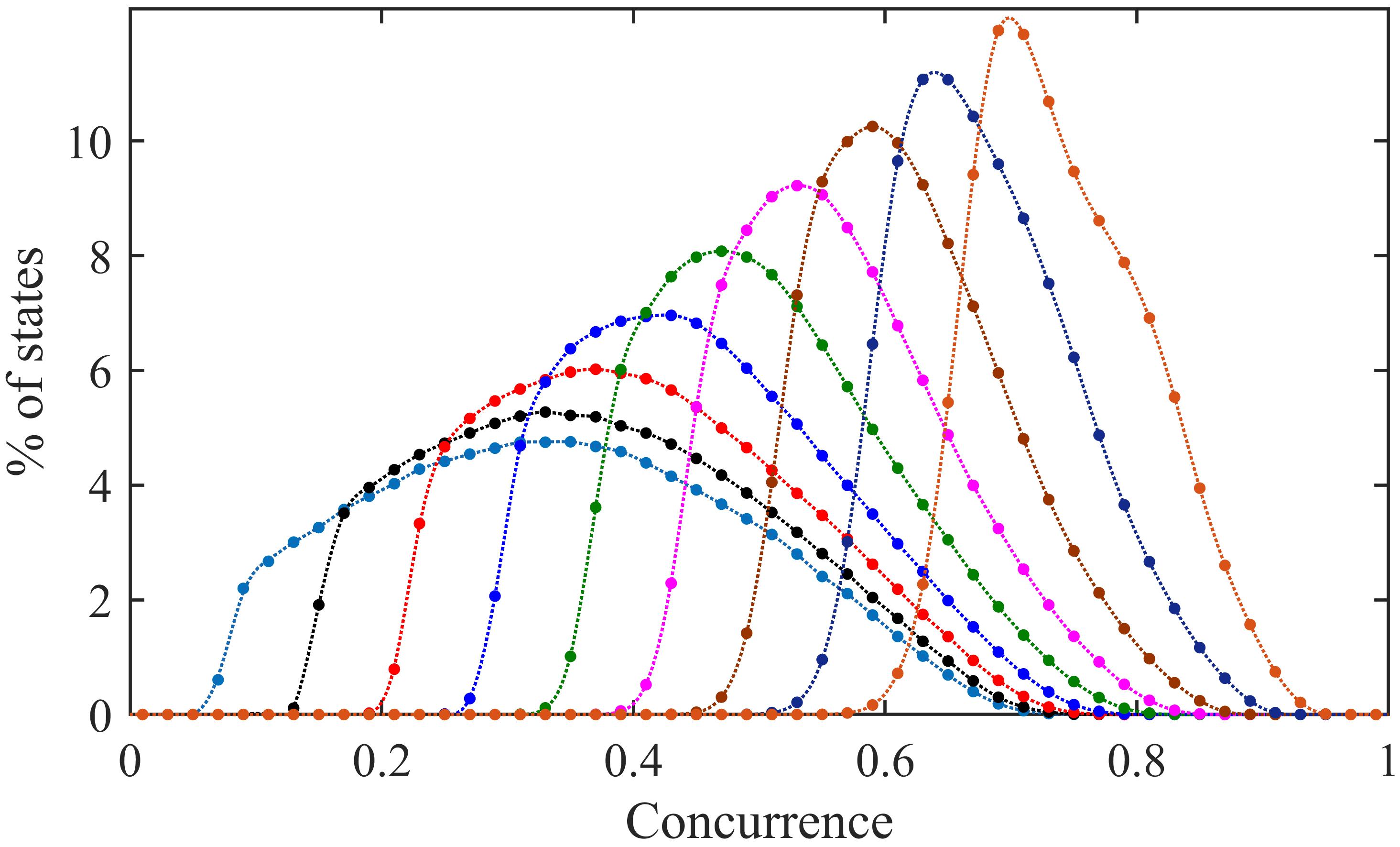}
    \caption{Spread of two-party pure quantum state entanglement in response to disorder in a single parameter. We plot here the relative frequency percentages  of  Haar-uniformly chosen two-qubit pure states \(|\psi\rangle\), given by Eq.~(\ref{e10}), with concurrence values in ebits in the ranges 0.1$\pm$0.01 (cyan dots), 0.2$\pm$0.01 (black dots), 0.3$\pm$0.01 (red dots), 0.4$\pm$0.01 (blue dots), 0.5$\pm$0.01 (green dots), 0.6$\pm$0.01 (magenta dots), 0.7$\pm$0.01 (brown dots), 0.8$\pm$0.01 (deep blue dots), 0.9$\pm$0.01 (orange dots), with disorder at $\alpha_1$ chosen from Gaussian distributions. These quantities, for the different cases are on the vertical axis, and are plotted against the concurrence values, with the latter ranging between $0$ and $1$, clustered in small sub-ranges. The 
    disorder averaged values plotted in the figure are for \(100\) disorder configurations. The figure does not change, up to the precision used, for averaging over \(50\) configurations. The precision is checked to three significant figures. The vertical axis is dimensionless, while the horizontal one is in ebits. The skewnesses of the plots are  0.15,  0.26, 0.35,  0.42,  0.47,  0.50,  0.49,  0.46,  0.36, respectively for the plots from left to right. Please see text for more details.
    }
  \label{res1}
\end{figure}

Before investigating the effects of disorders randomly drawn from non-Gaussian distributions, we will explore the reason behind the apparent non-intuitive effect of disorder on random quantum states, viz., the inhibition of the spread of entanglement in response to disorder. We focus on Haar uniformly chosen states with  values of concurrence, measured in ebits, in the range
\((C_i-\Delta,C_i+\Delta)\), separately for \(C_i = 0.1, 0.2, \ldots, 0.9\), and for \(\Delta = 0.01\).
%
These quantum states are perturbed at a single parameter ($\alpha_1$ of Eq.~(\ref{e10})) using disorder chosen from Gaussian distributions with $\mu_G=\alpha_1$, 
and $\gamma_G=1/2$. Fig.~\ref{res1} shows the response of this Gaussian disorder in a single parameter on the distribution of concurrence.
\textcolor{black}{All the curves are right skewed. The right skewness gradually increases with increase of \(C_i\) reaching a maximum at \(C_i=0.6\) and decreases thereafter.} 
{\textcolor{black}{The}} average value of entanglement, as obtained from these distributions, are exhibited in Table~\ref{number1}. 
{\color{black}{It}} seems that as we perturb a quantum state which has a certain value of concurrence, it has greater probability to transform into a state with higher or lower entanglement depending on whether the parent state had a value of entanglement that was lower or higher 
than 
the average entanglement of Haar uniformly distributed states in the ordered case.
The value of the latter is 0.589 ebits.
As an example, we may refer to the curve with black 
dots in Fig.~\ref{res1}, which corresponds to \(C_i=0.2\) ebits, which is therefore less than 0.589. After the insertion of disorder, the 
average of the distribution corresponding to \(C_i=0.2\) moves towards 0.589. The opposite happens for \(C_i=0.8\). 
%
%
%
%
We will find that this skewed effect of disorder is similar, although more pronounced, when disorder is applied in four parameters of the Haar uniformly chosen random quantum states.

\begin{table}[htbp]
\fontsize{7.2}{8.5}\selectfont

\centering
\begin{tabular}{ |c|c|c|c|c|c|c|c|c|c|c| } 
 \hline
 $C_i$ & $0.1$ & $0.2$ & $0.3$ & $0.4$ & $0.5$ & $0.6$ & $0.7$ & $0.8$ & $0.9$ \\ 
 \hline
 Average & $0.342$ & $0.373$ & $0.414$ & $0.461$ & $0.512$ & $0.567$ & $0.624$ & $0.683$ & $0.743$ \\ 
 \hline
\end{tabular}
\caption{Comparison of concurrences before and after the introduction of disorder at one parameter. 
The single parameter in which disorder is introduced is \(\alpha_1\) of Eq.~(\ref{e10}).
The first row denotes the midpoint of the range \((C_i - \Delta,C_i + \Delta)\). The second row denotes the average of the concurrences of the disorder-induced distribution of states for that range. See text for more details.
%
%
%
}

\label{number1}
\end{table}

\subsection{Non-Gaussian disorder}
\label{durbar-ek}

For investigating the question of genericity, the first of the removals of specificity is effected by considering non-Gaussian disorder distributions. Therefore, the ``second step'' mentioned above is followed to separately introduce disorder at $\alpha_1$ from the uniform and Cauchy-Lorentz distributions, instead of the Gaussian one. For the case of the uniform distribution, random numbers are selected from a uniform distribution with $\mu_U = \alpha_1$, where $\alpha_1$ is the random number generated in the first step, and $\gamma_U = 1/2$. In the Cauchy-Lorentz case, 
disorder is  introduced at $\alpha_1$ by choosing random numbers from a Cauchy-Lorentz distribution with median $x_0=\alpha_1$ and semi-interquartile range $\gamma_{C-L}=1/2$.  Note that the mean is equal to the median for the Gaussian and uniform distributions. The data for the uniform distribution is plotted as green crosses, and that for the Cauchy-Lorentz as blue pluses, in Fig.~\ref{f1}. 


As already reported above, the average entanglement for the Haar-uniformly chosen random pure two qubit states is $0.589$, while the corresponding standard deviation 
is $0.230$, in the ordered (clean) case. 
All numbers in the calculations for the two-qubit cases are correct to three significant figures. The equality symbols used in the corresponding figures are up to this precision. 
It can be seen from Fig.~\ref{f1} that introduction of disorder in the parameter $\alpha_1$ from Gaussian and uniform distributions  {\color{black}{slightly reduce}} the 
disorder averaged entanglement (equalling \(0.569\) in the former case, as already reported above, and \(0.581\) in the latter one). However, the 
disorder averaged standard deviations of the corresponding entanglement distributions are \textcolor{black}{significantly} reduced (being \(0.153\) in the case of  Gaussian disorder, as already reported above, and \(0.157\) in the uniform case). There are appreciable changes in both  mean and standard deviation (being respectively \(0.513\) and \(0.141\)) of the entanglement distribution when the disorder in the parameter $\alpha_1$ is from the Cauchy-Lorentz distribution.

The mean and the standard deviation provide important information about the relative frequency distribution of random states as a function of concurrence. The mean and standard deviation are respectively the first raw moment and the second central moment of a distribution. There are of course higher moments that can be analyzed to gather further information about the distribution under consideration.  To investigate the distribution further we employ two higher order central moments scaled by the standard deviation, viz. skewness (s) and kurtosis (k).  
We find that the plots in the ordered cases are left skewed, and after inclusion of disorder, the left-skewness  {\color{black}{increases}} slightly in all the three cases of disorder considered. 
The kurtosis increases when the disorders are introduced from the \textcolor{black}{Gaussian, uniform and} Cauchy-Lorentz distributions, indicating an increase of outliers. The values of skewness and kurtosis for the frequency distributions are mentioned in the caption of Fig.~\ref{f1}.

So we find that altering the distribution of disorder does not alter the fact that the disorder results in an inhibition of spread of the entanglement distribution. In fact, moving over to the Cauchy-Lorentz case makes the inhibition 
\textcolor{black}{a bit } more pronounced.

\subsection{Noisy two-qubit states}
\label{durbar-dui}

For two-party pure states, the local von Neumann entropy is a ``good'' measure of entanglement~\cite{PhysRevA.53.2046}
. And for two-qubit pure states, this is equivalent to the concurrence~\cite{PhysRevLett.78.5022,PhysRevLett.80.2245}. 
We however use concurrence to measure entanglement of two-qubit pure states, to be able to consider the response of disorder on entanglement of two-qubit pure states admixed with noise, thus leading to mixed states, for which, von Neumann entropy of local density matrices does not quantify entanglement, while concurrence does. 

The intention in this subsection is to look at the effect, on the response to disorder that we have already seen above, of noise admixture in the states involved. More precisely, we consider a Haar-uniformly generated set of states, \(|\psi\rangle\), given by Eq.~(\ref{e10}), with each admixed with white noise, so that the actual state is 
\begin{equation}
\varrho = p|\psi\rangle \langle \psi| + (1-p) \frac{1}{4} I_4.
\end{equation}
We consider a fixed \(p\) for every \(|\psi\rangle\), so that we can assume that we are dealing with a situation where any of the randomly generated states are passing through a fixed noisy channel, that admixes the input with white noise with a fixed noise level. The state \(\varrho\) of course depends on the state \(|\psi\rangle\) and the noise parameter \(p\), which are kept silent in the notation. Here, \(I_4\) represents the identity operator on the two-qubit Hilbert space.

\begin{figure}[h]
  \centering
    \includegraphics[width=\linewidth]{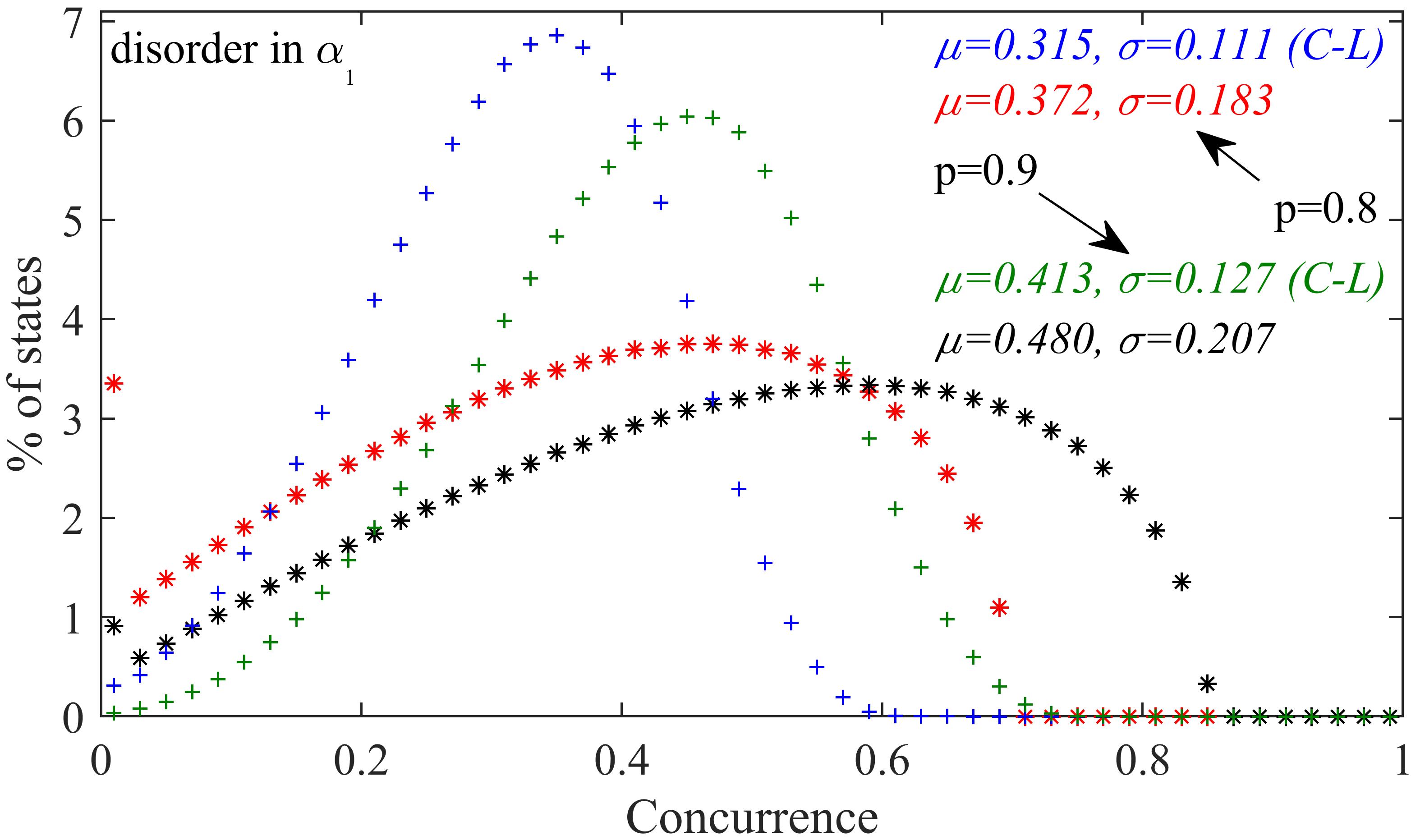}
    \caption{Effect of noise on the inhibited spread of two-party entanglement due to disorder insertion. The considerations in this figure are the same as in the preceding one, except that only the Cauchy-Lorentz disorder is considered, and that the states are admixed with white noise. The symbols are also different in this figure as compared to the preceding one. The black asterisks and the green pluses depict the case when noise-level is given by \(p=0.9\), with the asterisks being for the clean case and the pluses for the 
    disorder averaged situation. The red asterisks and the blue pluses are similarly for \(p=0.8\). For $p=0.9$ the skewness and kurtosis for the clean case are $s(p=0.9)=-0.288$ and $k(p=0.9)=2.19$, while for the disordered case the corresponding values are $s(C-L; p=0.9)=-0.313$ and $k(C-L; p=0.9)=2.66$. Similarly for $p=0.8$ the skewness and kurtosis for the clean case are $s(p=0.8)=-0.253$ and $k(p=0.8)=2.11$, while for the disordered case the corresponding values are $s(C-L; p=0.8)=-0.279$ and $k(C-L; p=0.8)=2.59$. Note that the `disordered distributions' have marginally greater left-skewness and marginally higher kurtosis. For a discussion on the additional non-monotonicity near zero entanglement that occurs in this noisy case for the curves corresponding to the clean situations, see text.
    }
  \label{natun-ek}
\end{figure}

We fix attention on the Cauchy-Lorentz disorder distribution, and compare the corresponding 
disorder averaged entanglement spread in the noisy states with that in the noisy case without disorder. We find that the feature of inhibition of spread remains valid even in this noisy situation. The results are plotted in Fig.~\ref{natun-ek}, where two different values of the noise level, viz. \(p=0.9\) and \(p=0.8\),  are considered. Both the mean and standard deviation are affected by the disorder, and in particular for \(p=0.9\), the 
disorder averaged standard deviation is 0.127, while the standard deviation in the clean case is 0.207. A similar inhibition of the spread is observed for \(p=0.8\), as exhibited in the same figure. 

Two points could be mentioned before we move away from this noisy case. Firstly, note that noise itself also leads to an inhibition of the spread. 

Secondly, the curves of the percentages of states, when plotted against entanglement (as quantified by concurrence) are non-monotonic. However, in the noiseless cases, all of them have a bell-like shape, and in particular do not  have any non-monotonicity near zero entanglement. In the noisy case, there appears an additional non-monotonicity near zero entanglement (in the clean cases).
This is not due to any numerical convergence problem, and, e.g., 
the percentage of zero entanglement states in the clean case for \(p=0.8\) is \(2.32\%\), which remains the same irrespective of whether \(10^7\) or \(10^6\) states are considered in the Haar-uniform generation, up to three significant figures.
This additional non-monotonicity near zero entanglement can be expected from the fact that the noise inflicted has brought the original set of states closer to the separable ball, and has resulted in more states clustered near the zero-entanglement point. We will come back to this point again when we consider noisy three-qubit states below, where this additional non-monotonicity will be absent. 
Note that the percentage of zero entanglement states decreases with increasing \(p\), as expected. We remember that increasing \(p\) corresponds to decrease in noise. Note also that this additional non-monotonicity does not remain in the disordered case, due to the smearing-out effect of the 
disorder averaging process.

\subsection{Disorder in multiple parameters}
\label{durbar-tin}

We revert to pure states, but now consider disorder in multiple parameters. Note that in the considerations until now, we have analyzed situations where there is disorder only in one parameter of the Haar-uniformly generated state, \(|\psi\rangle\), in Eq.~(\ref{e10}). 

\subsubsection{Two parameters}

\begin{figure}[h]
  \centering
    \includegraphics[width=\linewidth]{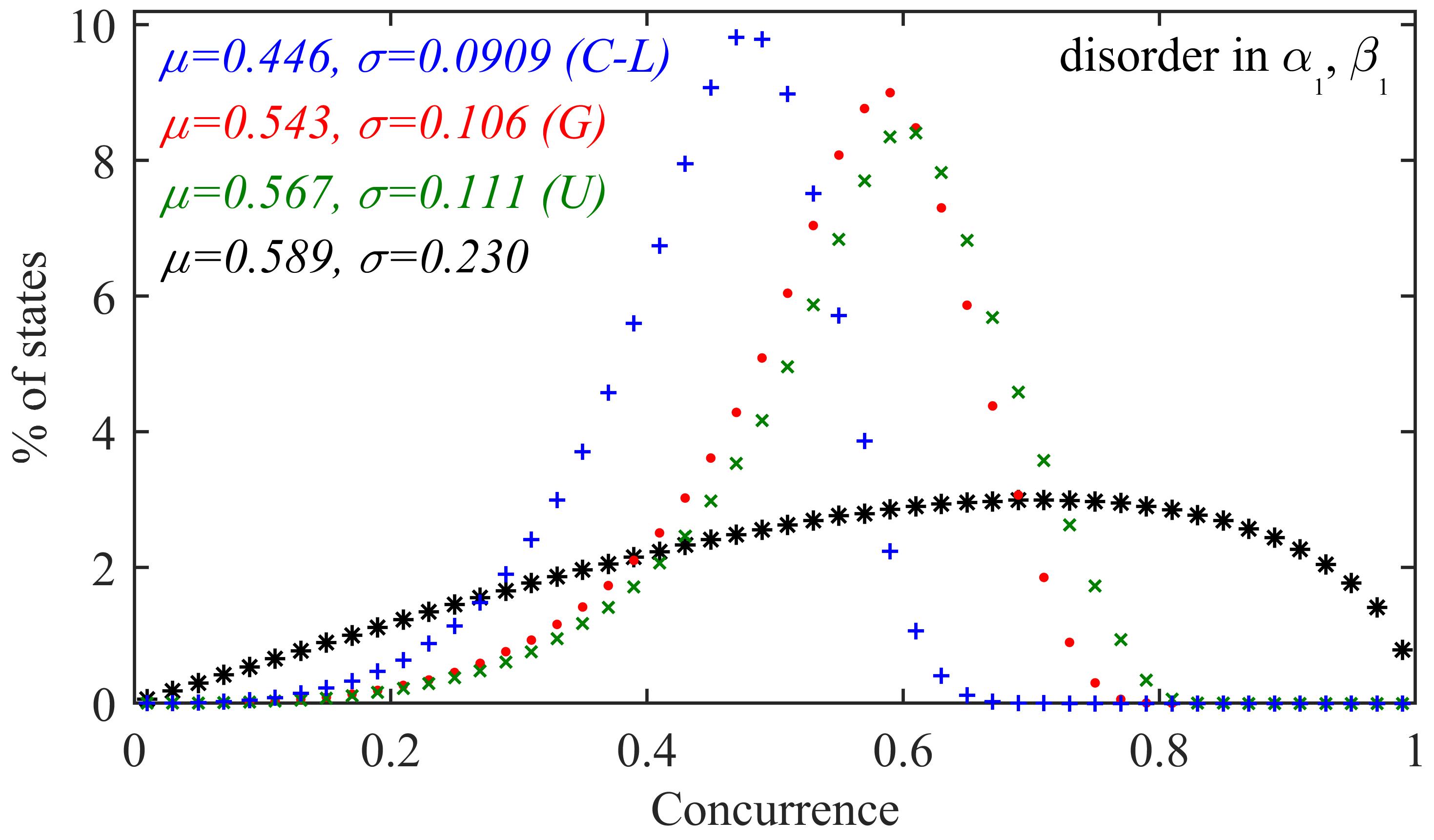}
    \caption{Further hindrance to entanglement spread as we introduce disorder in more parameters. The case of introducing disorder in two parameters is considered in this figure, while the case of doing the same for four parameters is considered in the succeeding figure.  The considerations are the same as in Fig.~\ref{f1}, except that the disorder is introduced in \(\alpha_1\) and \(\beta_1\), instead of just \(\alpha_1\). The skewness and kurtosis of the ordered plot are $s=-0.292$ and $k=2.20$, while for the plots in the disordered cases, the skewness and kurtosis are $s(U)=-0.770$, $k(U)=3.73$; $s(G)=-0.856$, $k(G)=3.80$; $s(C-L)=-0.729$, $k(C-L)=3.56$, for disorders from uniform, Gaussian, and Cauchy-Lorentz distributions respectively. The disordered plots are more left skewed in comparison with the clean case. 
    The kurtosis is seen to increase for the disordered cases. 
    }
  \label{f2}
\end{figure}

We now analyze the situation where disorder is 
introduced
in \emph{two} parameters, viz.  $\alpha_1$ and $\beta_1$, independently from Gaussian, uniform, or Cauchy-Lorentz distributions. Random numbers are selected from Gaussian, uniform distributions with $\mu_{G/U}=\alpha_1$, $\gamma_{G/U}=1/2$ and $\mu_{G/U}=\beta_1$, $\gamma_{G/U}=1/2$, while random numbers are chosen from Cauchy-Lorentz distributions with $x_0=\alpha_1$, $\gamma_{C-L}=1/2$ and $x_0=\beta_1$, $\gamma_{C-L}=1/2$. These random numbers are the new $\alpha_1$s and $\beta_1$s in Eq.~(\ref{e10}), while the other random numbers are the ones selected in the first step, post normalization. One hundred such pure states are generated and the average entanglement with disorder is calculated. The resulting entanglement distributions are plotted in Fig.~\ref{f2}.
It can be seen from Fig.~\ref{f2} that introduction of disorder in both the parameters $\alpha_1$ and $\beta_1$ from Gaussian and uniform distributions \textcolor{black}{reduces} 
the average entanglement {\textcolor{black}{slightly}} (being \(0.543\) in the Gaussian case and \(0.567\) in the uniform one). However, the standard deviations of the corresponding entanglement distributions are reduced further (being \(0.106\) in the Gaussian case and \(0.111\) in the uniform one) in comparison with the matching cases where disorder was introduced in the coefficient $\alpha_1$ only. There are appreciable changes in both  mean and standard deviation (being respectively \(0.446\) and \(0.0909\))  
of the entanglement distributions when the disorders in the parameters $\alpha_1$ and $\beta_1$ are randomly chosen from the Cauchy-Lorentz distribution, and again the standard deviation is lower than when a Cauchy-Lorentz disorder was inserted only in the parameter \(\alpha_1\) (compare with Fig.~\ref{f1}).

\subsubsection{Four parameters}

We have already seen that introducing disorder in two parameters leads to increased restriction, in comparison to when we introduce the same in one parameter, on the spread of entanglement on the span \([0,1]\). Does this increase in restriction to spread continue when we introduce disorder in even further parameters? In an attempt to find this answer, we now consider the effect of introducing disorder in four parameters. 

\begin{figure}[h]
  \centering
    \includegraphics[width=\linewidth]{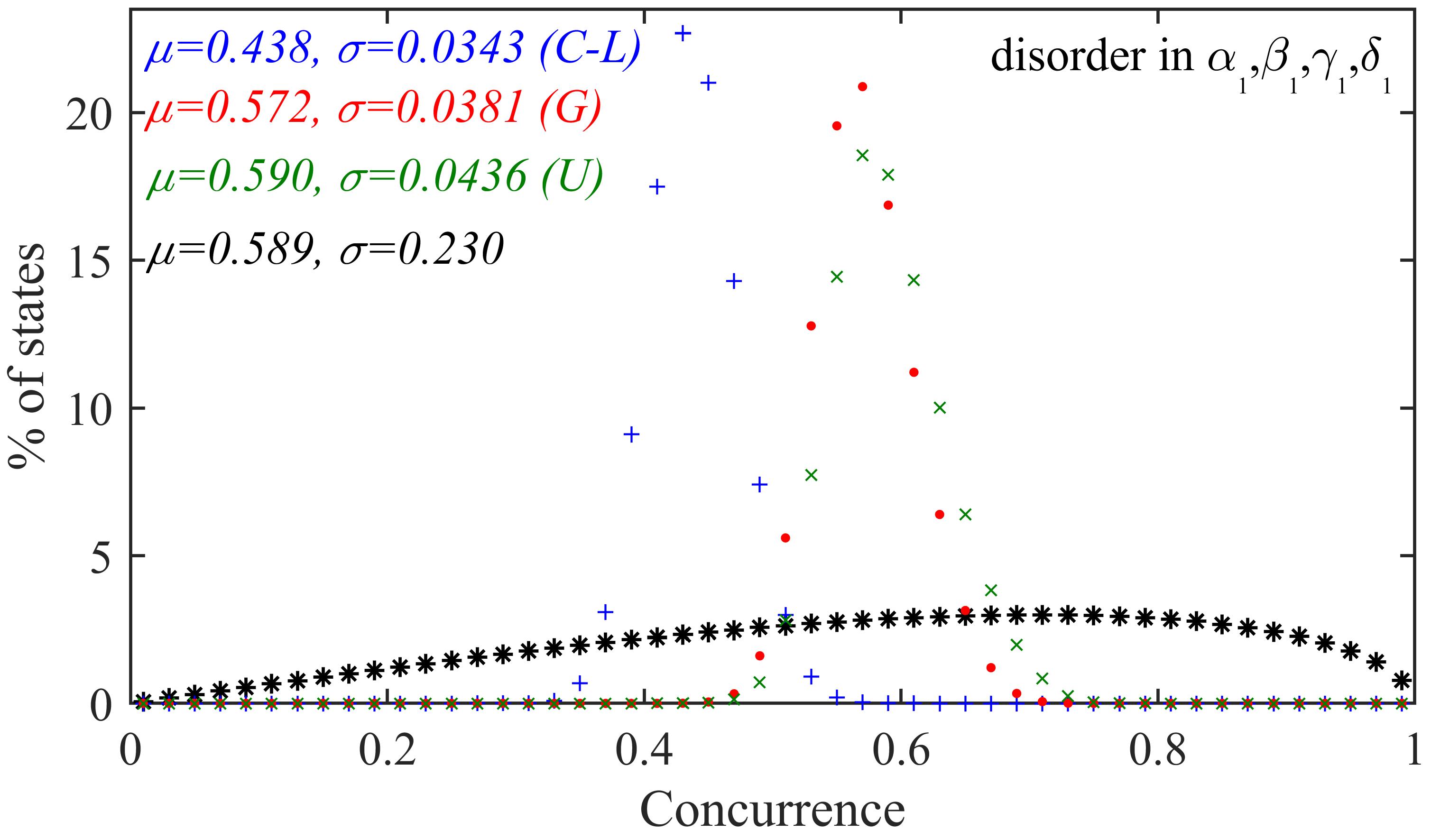}
    \caption{Disorder in four parameters lead to increased restriction, in comparison to the case when disorder was introduced in two parameters,  in spread of entanglement in two-qubit states. All considerations except the increased number of parameters is the same as in the preceding figure. The skewness and kurtosis of the ordered plot are $s=-0.292$ and $k=2.20$ as mentioned in figure captions before. For the disordered case,  the skewness and kurtosis are $s(U)=0.400$, $k(U)=3.05$; $s(G)=0.283$, $k(G)=3.02$; $s(C-L)=0.163$, $k(C-L)=3.00$, for disorders from uniform, Gaussian and Cauchy-Lorentz distributions respectively. The disordered distributions are now right skewed 
    \textcolor{black}{and the kurtosis is seen to increase}.
    }
  \label{f3}
\end{figure}

Therefore, we now 
introduce 
disorder in the parameters $\alpha_1$, $\beta_1$, $\gamma_1$, and $\delta_1$ independently from Gaussian, uniform, or Cauchy-Lorentz distributions. Random numbers for introduction of disorder are selected from Gaussian or uniform distributions with \begin{eqnarray*}
\mu_{G/U}=\alpha_1, \gamma_{G/U}=1/2,\\ \mu_{G/U}=\beta_1, \gamma_{G/U}=1/2,\\ \mu_{G/U}=\gamma_1, \gamma_{G/U}=1/2,\\ \mu_{G/U}=\delta_1, \gamma_{G/U}=1/2,
\end{eqnarray*}
while random numbers are chosen from Cauchy-Lorentz distributions with \begin{eqnarray*}
x_0=\alpha_1, \gamma_{C-L}=1/2,\\ x_0=\beta_1, \gamma_{C-L}=1/2,\\ x_0=\gamma_1, \gamma_{C-L}=1/2,\\ x_0=\delta_1, \gamma_{C-L}=1/2.
\end{eqnarray*}
These random numbers are the new $\alpha_1$s, $\beta_1$s, $\gamma_1$s, and $\delta_1$s, while the other random numbers are the ones selected in the first step
\textcolor{black}{post} normalization. Once more, one hundred such pure states are generated and the 
disorder averaged entanglement is calculated. The resulting entanglement distributions are plotted in Fig.~\ref{f3}.
It can be seen from Fig.~\ref{f3} that introduction of disorder in the four parameters $\alpha_1$, $\beta_1$, $\gamma_1$, and $\delta_1$ from Gaussian or uniform distributions 
does not appreciably affect the average entanglement (being \(0.572\) in the Gaussian case and \(0.590\) in the uniform case) with respect to its value in the clean case. However, the standard deviations of the corresponding entanglement distributions are \textcolor{black}{significantly} reduced further than their values in the case when disorder was introduced in two parameters, which themselves were lower than the clean case value. Specifically, the standard deviations are \(0.0381\) and \(0.0436\), respectively for introduction of Gaussian and uniform disorders.
There are considerable decreases in both the mean and standard deviation (respectively, \(0.438\) and \(0.0343\)) of the entanglement distributions when the disorder in the parameters $\alpha_1$, $\beta_1$, $\gamma_1$, and $\delta_1$ are randomly and independently chosen from the Cauchy-Lorentz distribution.

\begin{figure}[h]
  \centering
    \includegraphics[width=\linewidth]{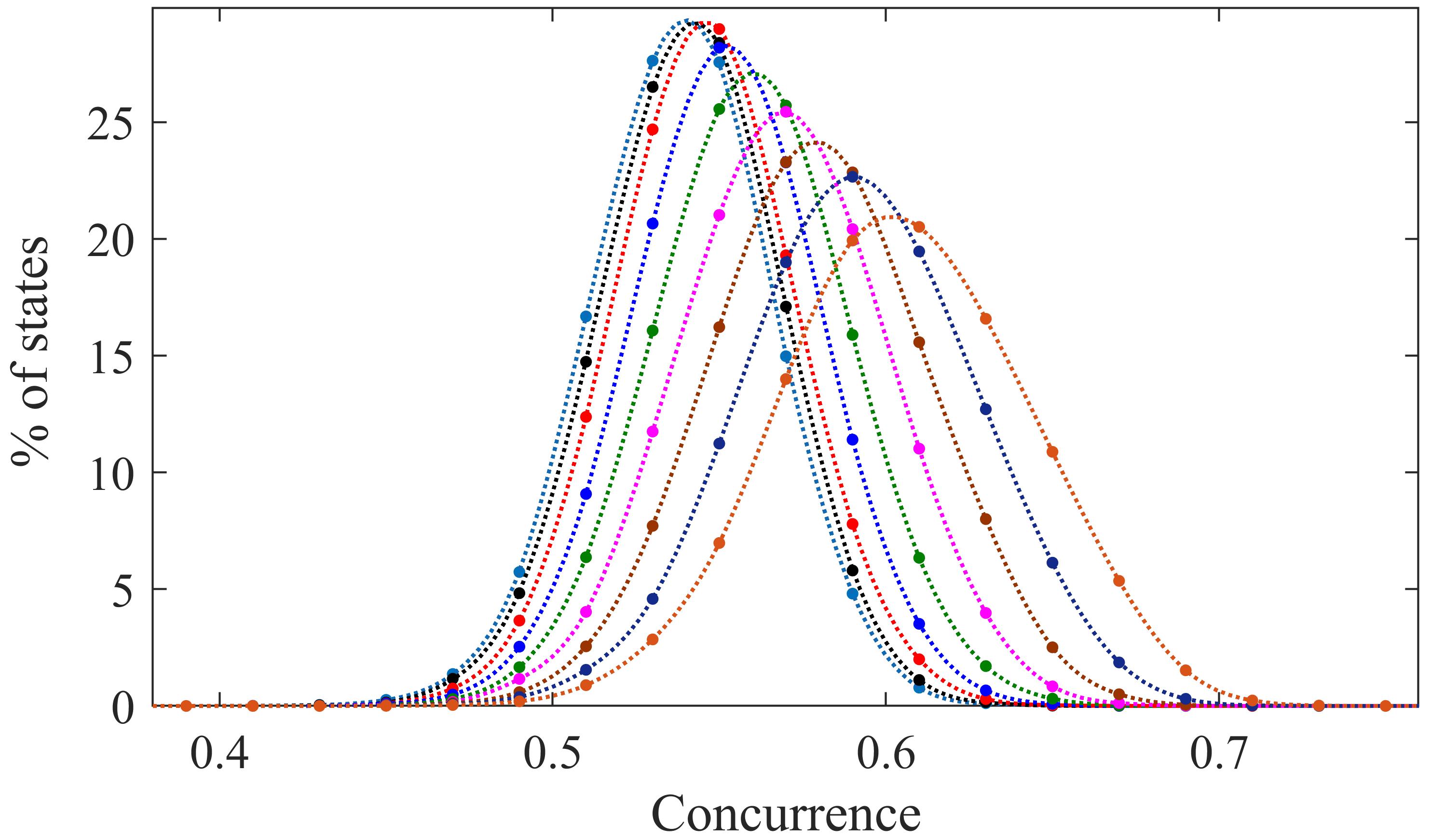}
    \caption{Spread of two-party pure quantum state entanglement in response to disorder in four parameters.
    The considerations are the same as in Fig.~\ref{res1}, except that the 
%
 disorder is introduced in four state parameters.
 The skewnesses of the plots are  -0.075, -0.066,  -0.068,  -0.035,  -0.004,  -0.029,  -0.030,  -0.056,  -0.057, from left to right. See text for more details.} 
  \label{res2}
\end{figure}

Let us now revisit our efforts in finding the reason behind the inhibition of the spread of entanglement due to disorder. We had pointed to a possible reason at the beginning of this section, around Fig.~\ref{res1} and Table~\ref{number1}. We again focus on Haar uniformly chosen states with  values of concurrence measured in ebits in the ranges 0.1$\pm$0.01, 0.2$\pm$0.01, \(\ldots\), 0.9$\pm$0.01.  These quantum states are perturbed at four parameters $\alpha_1$, $\beta_1$, $\gamma_1$, and  $\delta_1$  using disorder chosen from Gaussian distributions.
Fig.~\ref{res2} depicts the effect of this  disorder insertion  for the chosen sets of random states. 
\textcolor{black}{\textcolor{black}{All the curves are slightly left skewed. The numerical values of skewness gradually increase with increase of \(C_i\), reaching a maximum for the curve corresponding to \(C_i=0.5\), and decreases thereafter.}} The curves themselves are significantly right-shifted for states with \(C_i=\) 0.1, 0.2, 0.3, 0.4, \textcolor{black}{0.5} and significantly left-shifted for states with  the same equalling \textcolor{black}{0.6, 0.7,} 0.8 and 0.9. Note that the average value of entanglement, shown in Table~\ref{number2} varies accordingly. 
The findings therefore again point to the same realization that 
as we perturb a quantum state which has a certain value of concurrence, it has greater probability to transform into a state with higher or lower entanglement depending on whether the parent state has an entanglement that is lower or higher than the average entanglement in the ordered case.
We moreover find here that the effect of disorder is more pronounced in this case in comparison with the case when disorder is applied in only one parameter of the Haar uniformly chosen random quantum states.

\begin{table}[htbp]
\fontsize{7.2}{8.5}\selectfont
\centering
\begin{tabular}{ |c|c|c|c|c|c|c|c|c|c|c| } 
 \hline
 $C_i$ & $0.1$ & $0.2$ & $0.3$ & $0.4$ & $0.5$ & $0.6$ & $0.7$ & $0.8$ & $0.9$ \\ 
 \hline
 Average & $0.538$ & $0.541$ & $0.546$ & $0.552$ & $0.560$ & $0.569$ & $0.580$ & $0.592$ & $0.605$ \\ 
 \hline
\end{tabular}
\caption{Comparison of concurrences before and after the introduction of disorder at four state parameters.
The considerations are the same as in Table~\ref{number1}, except for the number of parameters in which disorder is introduced. 
}
\label{number2}
\end{table}

\subsection{Effect of variation in dispersion of disorder}
\label{durbar-char}

\begin{figure}[h]
  \centering
    \includegraphics[width=\linewidth]{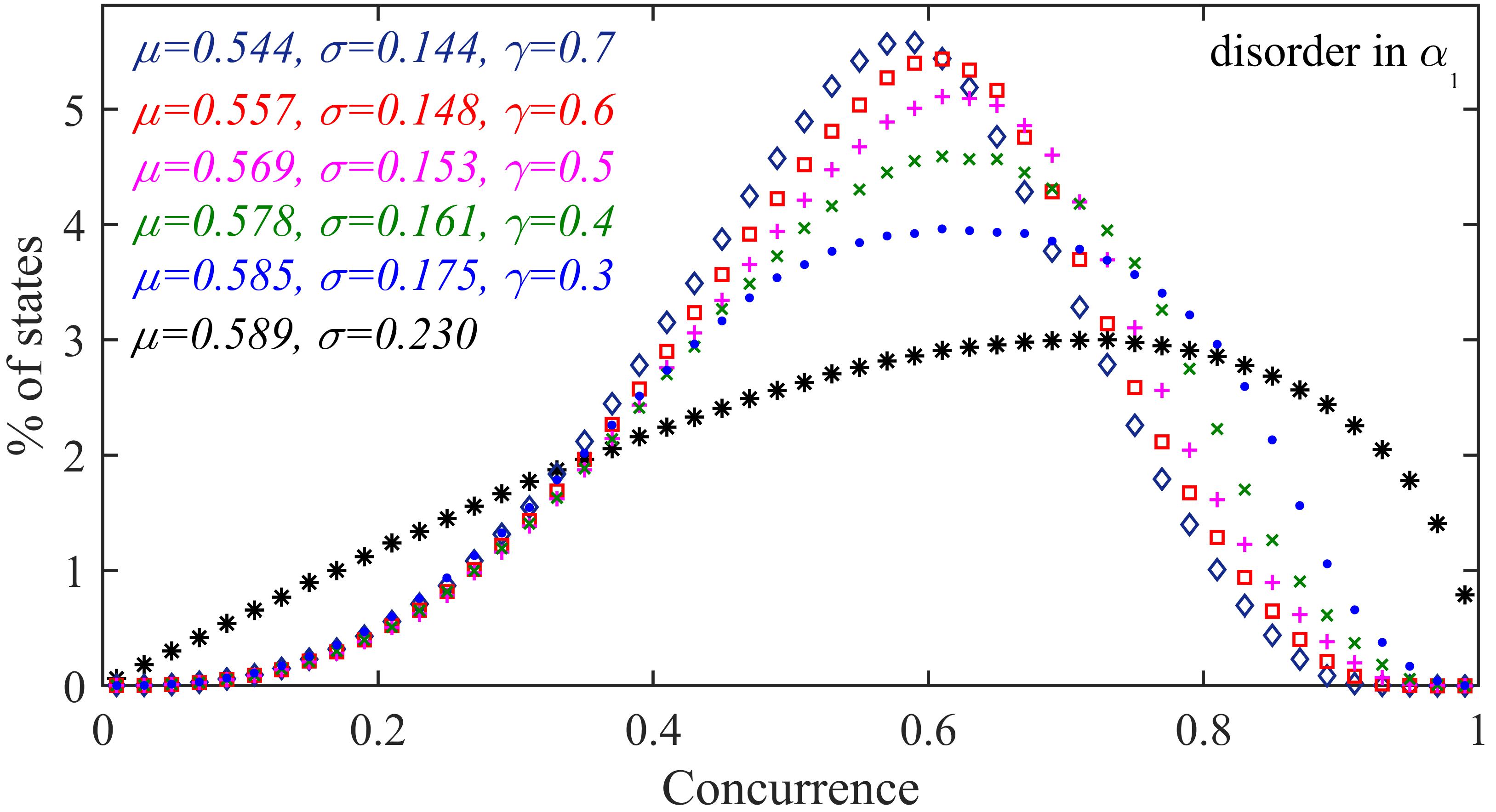}
    \caption{Effect of variation of strength of disorder. The considerations are the same as in Fig.~\ref{f1}, except that only the Gaussian disorder is considered, and that different semi-interquartile ranges are used for the plots. The symbols are also different here. Black asterisks represent the case when there is no disorder, and has appeared in several figures before, including Fig.~\ref{f1}. The curve with pink pluses is for \(\gamma_G = 1/2\), and was also present in Fig.~\ref{f1}. The blue (cyan) dots, green crosses, red squares, and blue-black diamonds are respectively for \(\gamma_G = 0.3\), \(0.4\), \(0.6\), and \(0.7\). Note that the suffix of \(\gamma_G\) is kept silent in the figure legend on the plot. 
    }
  \label{varying-gamma}
\end{figure}

We revert to disorder in a single parameter, but stay with pure states.
The dispersion, as quantified by the semi-interquartile range, has until now been kept fixed at \(1/2\), and which we wish to vary now to see its effect on the inhibition of spread of entanglement. Changing the semi-interquartile range can be interpreted as varying the strength of the disorder introduced, with increase of semi-interquartile range implying increase of the strength.
We fix attention on the Gaussian disorder for this purpose, and also introduce the disorder only on a single parameter. 
Random numbers are selected from Gaussian distributions with $\mu_{G}=\alpha_1$, and  $\gamma_{G}$ varying from \(0.3\) to \(0.7\) at intervals of \(0.1\). These random numbers are the new $\alpha_1$s in Eq.~(\ref{e10}), while the other random numbers are the ones selected in the first step,
\textcolor{black}{post} normalization. Just like in the other cases, one hundred such pure states are generated with the disorder being chosen from the Gaussian distribution, and then the average entanglement is calculated. The resulting entanglement distributions (for different semi-interquartile ranges) are plotted in Fig.~\ref{varying-gamma}.
It can be seen from Fig.~\ref{varying-gamma} that the average entanglement \textcolor{black}{decreases minimally} 
with the increase of $\gamma_G$, while the standard deviations of the entanglement distributions \textcolor{black}{significantly} decrease with the increase of $\gamma_G$. Therefore, as the strength of disorder in the states increases, the standard deviations of the resulting entanglement distributions decrease. This feature can help us understand the reason why the Cauchy-Lorentz distribution has consistently been found to lead to greater inhibition of the spread of entanglement in the previous cases, whenever compared with Gaussian and uniform disorder distributions with same semi-interquartile ranges. The Gaussian and uniform distributions have a finite mean, unlike the Cauchy-Lorentz distribution. 
For two probability distributions (among Gaussian, uniform, and Cauchy-Lorentz) having the same semi-interquartile range, but with one having a finite mean and the other without, we can say that the latter has a 
longer ``reach'' in its domain of definition, viz. the real line, that has led to the non-existence of the mean (\(\int_{-\infty}^{+\infty} f(x) dx\) exists and is finite, but \(\int_{-\infty}^{+\infty} x f(x) dx\) does not exist, for a probability density \(f(x)\) on the real line). And consequently, we can interpret that the latter has a higher dispersion (spread), even though it has the same semi-interquartile range as the former. 

\subsection{Three-qubit pure states}
\label{durbar-panch}

\begin{figure}[h]
  \centering
    \includegraphics[width=\linewidth]{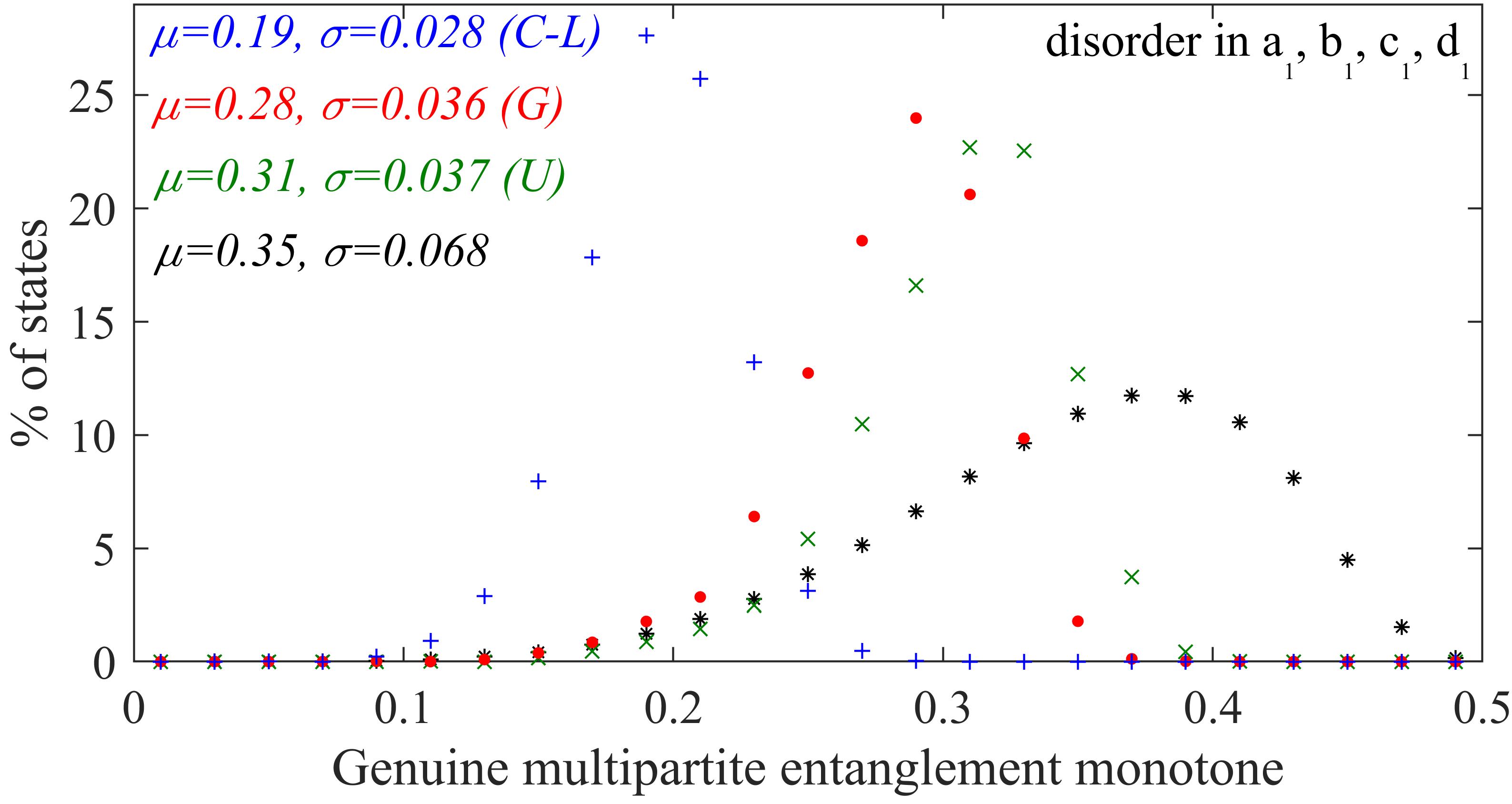}
    \caption{Inhibition of spread of entanglement in the pure three-qubit state space due to insertion of disorder in state parameters. We plot here the relative frequency percentages of Haar-uniformly generated random three-qubit pure states (with and without disorder at $a_1$, $b_1$, $c_1$, $d_1$) against the JMG entanglement monotone, with the latter ranging 
    between $0$ and $1/2$. The black asterisks correspond to percentage of random three-qubit pure states generated Haar uniformly, while the other three curves depict entanglement distributions for random three-qubit pure states with disorder at $a_1$, $b_1$, $c_1$, $d_1$ chosen from Gaussian (\(G\), red dots), uniform (\(U\), green crosses), or Cauchy-Lorentz (\(C-L\), blue pluses) distributions.
    The numbers in this figure and the succeeding one are correct to two significant figures. The window of the entanglement monotone used for calculating the percentages is 0.02. Both axes represent dimensionless quantities. The skewness and kurtosis of the ordered plot are $s=-0.55$ and $k=2.9$. For the disordered cases, the skewness and kurtosis are $s(U)=-0.83$, $k(U)=4.1$; $s(G)=-0.78$, $k(G)=3.9$; $s(C-L)=-0.34$, $k(C-L)=3.3$ for disorders from uniform, Gaussian, and Cauchy-Lorentz distributions respectively. 
    {\textcolor{black}{The left skewness of the frequency distributions, with disorders introduced from the uniform and Gaussian distributions, are seen to increase compared to the clean case. However, the left skewness of the frequency distribution, with disorders introduced from the Cauchy-Lorentz distribution, decreases compared to the clean case. The kurtosis of the disorder induced frequency distributions increase in all the cases, indicating an increase of outliers.}} 
    }
    
  \label{f4}
\end{figure}

We now move over to the three-qubit case, considering pure states in this subsection. The succeeding subsection deals with noisy (mixed) three-qubit states. 
%
The entanglement measure considered in both this and the succeeding sections is the JMG 
entanglement monotone.
A three-qubit random pure state can be represented as
\begin{equation} \label{three-qubit-state}
\begin{split}
 \ket{\Psi}=&(a_1+ia_2)\ket{000}+(b_1+ib_2)\ket{001}\\
 &+(c_1+ic_2)\ket{010}+(d_1+id_2)\ket{011}\\
 &+(e_1+ie_2)\ket{100}+(f_1+if_2)\ket{101}\\
 &+(g_1+ig_2)\ket{110}+(h_1+ih_2)\ket{111}.
 \end{split}
\end{equation}
where 
 \(a_1\), \(a_2\), \(b_1\), \(b_2\), \(c_1\), \(c_2\), \(d_1\), \(d_2\), \(e_1\), \(e_2\), \(f_1\), \(f_2\), \(g_1\), \(g_2\), \(h_1\), \(h_2\) are real numbers, constrained by the normalization condition, \(\langle \Psi|\Psi\rangle =1\). 
 The state in 
 Eq.~(\ref{three-qubit-state}) can be  Haar-uniformly generated by randomly choosing the sixteen real coefficients from independent  normal distributions of vanishing mean 
 and unit standard deviation,
 followed by a  normalization.
A large number ($2\times10^4$) of such random pure states are generated, normalized, and the entanglement monotone of each state is calculated. Next, we introduce disorder at $a_1$, $b_1$, $c_1$, and $d_1$, independently from Gaussian, uniform, or  Cauchy-Lorentz distribution functions. The disorder is introduced 
by choosing random numbers from Gaussian and uniform  distributions with
\begin{eqnarray*}
\mu_{G/U} = a_1, \gamma_{G/U} = 1/2,\\ \mu_{G/U} = b_1, \gamma_{G/U} = 1/2,\\ \mu_{G/U} = c_1, \gamma_{G/U} = 1/2,\\  \mu_{G/U} = d_1, \gamma_{G/U} = 1/2,
\end{eqnarray*}
where $a_1$, $b_1$, $c_1$, and $d_1$ are the random numbers generated in the first step. 
Cauchy-Lorentz distributions with 
\begin{eqnarray*}
x_0=a_1, \gamma_{C-L}=1/2,\\ x_0=b_1, \gamma_{C-L}=1/2,\\ x_0=c_1, \gamma_{C-L}=1/2,\\ x_0=d_1, \gamma_{C-L}=1/2,
\end{eqnarray*}
are also used
to introduce disorders at $a_1$, $b_1$, $c_1$, and $d_1$. These random numbers are the new $a_1$, $b_1$, $c_1$, and $d_1$,  while the other random numbers are the ones selected in the first post normalization. For every set of values of $a_1$, $b_1$, $c_1$, and $d_1$, chosen in the first step, fifty disordered states are generated by employing the above mentioned procedure. The random pure states are normalized and their average entanglement (the 
disorder averaged JMG entanglement monotone) calculated. The resulting entanglement distributions are plotted in Fig.~\ref{f4}.
It can be seen from Fig.~\ref{f4} that the average entanglement of random three-qubit pure states chosen Haar uniformly is $0.35$, while the standard deviation of the entanglement distribution is $0.068$. 
The average entanglement (being \(0.31\), \(0.28\), and \(0.19\) for uniform, Gaussian, and Cauchy-Lorentz cases respectively) as well as the standard deviations (being \(0.037\), \(0.036\), and \(0.028\) for uniform, Gaussian, and Cauchy-Lorentz cases respectively) of the entanglement distributions are reduced when disorder is introduced in the coefficients $a_1$, $b_1$, $c_1$, and $d_1$ from the uniform, Gaussian, and Cauchy-Lorentz distributions. 



\subsection{Noisy three-qubit states}
\label{durbar-chhoi}

One could have used simpler measures to quantify genuine multiparty entanglement, if we were required to deal with only pure three-qubit states. An example of such a measure is the generalized geometric measure~\cite{PhysRevA.81.012308,de2010bound,PhysRevA.90.032301,PhysRevA.94.022336,https://doi.org/10.1111/j.1749-6632.1995.tb39008.x,PhysRevA.68.042307,PhysRevA.77.062304}.
 We have however chosen to work with the JMG entanglement monotone due to its tractability for mixed multiparty states, to be considered in this subsection. 

\begin{figure}[h]
  \centering
    \includegraphics[width=\linewidth]{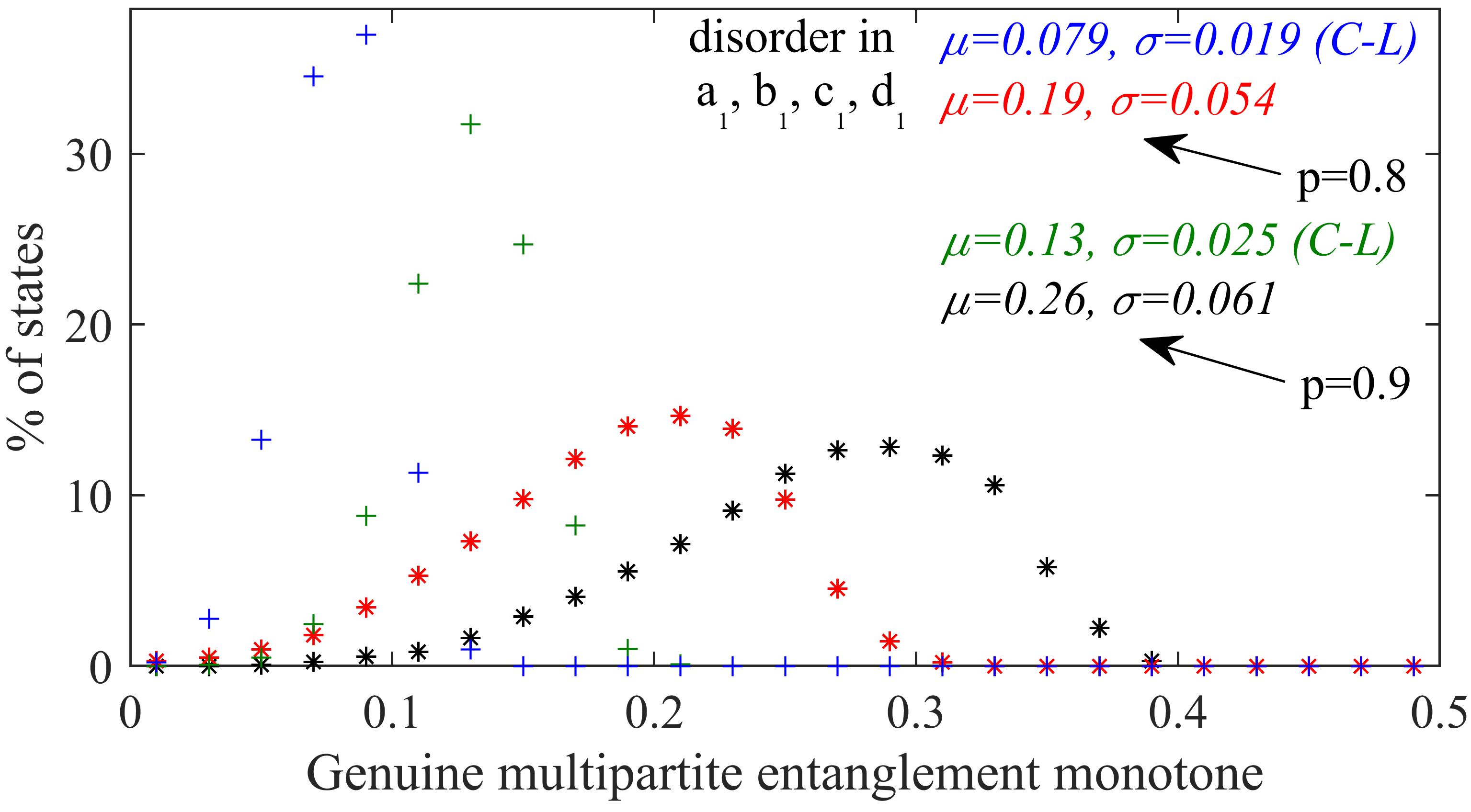}
    \caption{Further disorder-induced inhibition to spread of entanglement of three-qubit states in presence of noise. The considerations are exactly the same as in the preceding figure, except that the states are admixed with white noise, and that only Cauchy-Lorentz disorder is considered. The noise levels considered are given by \(p=0.9\) and \(p=0.8\). For \(p=0.9\), the black asterisks and green pluses represent the clean and disorder-averaged cases respectively. And for \(p=0.8\), the corresponding symbols are respectively red asterisks and blue pluses. For $p=0.9$, the skewness and kurtosis for the clean case are $s(p=0.9)=-0.55$ and $k(p=0.9)=2.9$, while for the disordered case, the corresponding values are $s(C-L; p=0.9)=-0.30$ and $k(C-L; p=0.9)=3.2$ respectively. For $p=0.8$, the skewness and kurtosis for the clean case are $s(p=0.8)=-0.50$ and $k(p=0.8)=2.9$, while for the disordered case, the corresponding values are $s(C-L; p=0.8)=-0.28$ and $k(C-L; p=0.8)=3.1$. The left skewness of the disordered plots decrease in comparison with the clean cases.
    The kurtosis of the disordered plots increase slightly in comparison with the clean cases, indicating a slightly increased percentage of outliers.}
  \label{natun-dui}
\end{figure}

Indeed, in this subsection, we wish to investigate the effect of noise on the restriction of spread of the JMG entanglement monotone in the space of three-qubit states. Precisely, we consider the state 
\begin{equation}
\tilde{\varrho} = p|\Psi\rangle \langle \Psi| + (1-p) \frac{1}{8}I_8,
\end{equation}
for every \(|\Psi\rangle\) (see Eq.~(\ref{three-qubit-state})) generated in the clean or the disordered cases. Here, \(I_8\) is the identity operator on the  three-qubit Hilbert space. This is exactly similar to the analysis in Sec.~\ref{durbar-dui}, except that we are considering three-qubit states now, and the measure is the JMG entanglement monotone. The effect obtained is again very similar, viz. noise leads to further inhibition of the spread of entanglement. The details of the numerical simulations are presented in Fig.~\ref{natun-dui}. We have focused attention only on the Cauchy-Lorentz disorder. Just like for the two-qubit case, noise restricts the spread of entanglement even in the clean cases. 
Let us also add that the additional non-monotonicity near zero-entanglement that we had observed in the two-qubit clean case, is absent here in the three-qubit one. We believe that the reason is that the volume of  separable states within the set of all quantum states (density matrices) decreases with increase in the number of parties~\cite[and references therein]{PhysRevA.58.883,PhysRevA.60.3496,PhysRevA.72.032304}. 

\section{Conclusion} \label{IV}
 We have analyzed the response to introduction of disorder in multiparty quantum state parameters on the entanglement in those states. We have considered two-qubit and three-qubit states in the analysis, with the entanglement measure considered for two-qubit states being the concurrence and that for three-qubit states being the Jungnitsch-Moroder-G{\" u}hne genuine multiparty entanglement monotone. The relative frequency percentages of states for given (small) windows of entanglement are not uniform for all ranges of the entanglements. We measured this non-uniformity by the standard deviation of the entanglement distribution of these percentages. We found that insertion of disorder in the state parameters generically shrinks the standard deviation from its clean-case value (i.e., value in the corresponding case without disorder). 
 
 We began with Gaussian disorder in a single state parameter for two-qubit pure states, and found that the distribution of the 
 disorder averaged entanglements has a lower standard deviation than the clean case. We then removed the specificities in this case in several ways. We considered non-Gaussian cases, viz. when disorder is introduced by using the uniform distribution as well as that using the  Cauchy-Lorentz one, where the latter one is different from the Gaussian and uniform distributions in that it does not have a mean. The Cauchy-Lorentz distribution turned out to be the one that provided the most hindrance to the spread of entanglement. We also considered the case when disorder is introduced in several state parameters, with more hindrance obtained as we increased the number of parameters in which disorder is inserted. We also found that increasing the strength of the disorder increases the localization effect on the entanglement spread.

 We also considered the response of disorder introduction in parameters of three-qubit pure states, and found that the inhibition of the spread of entanglement - genuine multiparty entanglement  in this case - can again be seen in this case. 
 For both two- and three-qubit cases, we considered the effect of noise on the phenomenon of inhibition of spread of entanglement in response to disorder introduction. 

Why does an entanglement measure that is allowed to span over a certain range does not cover that uniformly? The answer could be because the physical characteristic, viz. entanglement, is a nonlinear function of the state parameters. A peak develops in the allowed range of entanglement, when random quantum states are chosen. The distribution therefore has an average, and a nontrivial spread. It may intuitively seem that if we  perturb the state parameters, the distribution of the entanglement will also be perturbed: the mean will change and the spread will increase. What we saw is that while the mean does often change, the exact opposite happens for the spread: it decreases. Moreover, the stronger the perturbation, the stronger is the decrease in spread. 
This phenomenon can be understood by referring to how the behavior of  entanglement of a quantum state depends on its entanglement content, when perturbed. We performed this analysis for sets of two-qubit pure states having different amounts of entanglement - with finite precision - and when the disorder is inserted in a single state parameter and, separately, in four parameters.    
We observed that when perturbed, there is a large probability for a state with 
an entanglement lower (higher) than the ``average entanglement''  to jump to one with more (less) entanglement than in the parent state. Here, the average entanglement is the mean entanglement of Haar uniformly distributed states in the entire relevant Hilbert space.
%
%
%
And this leads to a clustering effect, driving the disorder-affected system to have a low dispersion in the entanglement distribution. We believe that the results will be of importance for considerations in quantum technologies as well as for understanding the black hole information paradox.


\begin{acknowledgments}
US acknowledges partial support from the Department of Science and Technology, Government of India through the QuEST  grant (grant number DST/ICPS/QUST/Theme-3/2019/120).
\end{acknowledgments}

\bibliography{2nd_paper}

\end{document}